\documentclass[12pt]{sensors}
\pdfoutput=1
\pagerange{4821-4850}
\doinum{10.3390/sensors}
\pubyear{2008}
\pubvolume{8}

\usepackage{amsmath}
\usepackage{amsfonts}
\usepackage{url}
\usepackage{algorithmic}
\usepackage{algorithm}
\usepackage{graphicx}

\newcommand{\norm}[1]{\ensuremath{\left\Vert #1 \right\Vert}}

\newcommand{\R}{\ensuremath{\mathbb{R}}}
\newcommand{\B}[1]{\ensuremath{\mathbf{#1}}}

\def\argmax{\mathop{\rm arg\,max}}

\arttype{Article}

\title{Distributed Principal Component Analysis for Wireless Sensor Networks}

\author{Yann-A\"el Le Borgne$^{1,\star}$, Sylvain Raybaud$^{2}$, and Gianluca Bontempi$^{1}$}

\address{
$^{1}$ Machine Learning Group, D\'epartement d'Informatique, Facult\'e des Sciences, Universit\'e Libre de Bruxelles, Boulevard du Triomphe, 1050 Brussels, Belgium\\
$^{2}$  \'Ecole Normale Sup\'erieure de Cachan, 61, Avenue du Pr\'esident Wilson, 94235 Cachan Cedex, France\\
E-mails: yleborgn@ulb.ac.be; sraybaud@dptmaths.ens-cachan.fr; gbonte@ulb.ac.be \\[12pt]
$^{\star}$ Author to whom correspondence should be addressed.\\[12pt]
{\em Received:  27 May 2008; in revised form: 29 July 2008 / Accepted: 4 August 2008 / \\Published: 11 August 2008}}

\abstract{

The Principal Component Analysis (PCA) is a data dimensionality reduction technique well-suited for processing data from sensor networks. It can be applied to tasks like compression, event detection, and event recognition. This technique is based on a linear transform where the sensor measurements are projected on a set of \emph{principal components}. When sensor measurements are correlated, a small set of principal components can explain most of the measurements variability. This allows to significantly decrease the amount of radio communication and of energy consumption. In this paper, we show that the power iteration method can be distributed in a sensor network in order to compute an approximation of the principal components. The proposed implementation relies on an aggregation service, which has recently been shown to provide a suitable framework for distributing the computation of a linear transform within a sensor network. We also extend this previous work by providing a detailed analysis of the computational, memory, and communication costs involved. A compression experiment involving real data validates the algorithm and illustrates the tradeoffs between accuracy and communication costs.
}


\begin{document}

\section{Introduction}

\label{sec:intro}

Efficient in-network data processing is a key factor for enabling
wireless sensor networks (WSN) to extract useful information and an
increasing amount of research has been devoted to the development of
data processing
techniques~\cite{kumar2002csa,intanagonwiwat2000dds,pattem2004isc,xiong2004dsc}.
Wireless sensors have limited resource constraints in terms of
energy, network data throughput and computational power. 
In particular, the radio communication is an energy consuming task and
is identified in many deployments as the primary factor of sensor
node's battery exhaustion \cite{akyildiz2002wsn}. Emitting or receiving a packet is indeed orders of magnitude more energy consuming than elementary computational operations. The reduction of the
amount of data transmissions has therefore been recognized as a
central issue in the design of wireless sensor networks data gathering
schemes \cite{ilyas2004hsn}. Data compression is often acceptable in
real settings since raw data collected by sensors typically contain a
high degree of spatio-temporal redundancies
\cite{krishnamachari2002ida,akyildiz2002wsn,heidemann2001bew,intanagonwiwat2002ind}. 
In fact, most applications only require approximated or high-level information,
such as the average temperature in a room, the humidity levels in a
field with a $\pm 10\%$ accuracy, or the detection and position of a
fire in a forest. 

An attractive framework for processing data within a sensor network is
provided by the data aggregation services such as those developed at
UC Berkeley (TinyDB and TAG projects)
\cite{madden2002tta,madden2005taq}, Cornell University (Cougar)
\cite{yao2002can}, or EPFL (Dozer)\cite{burri2007dul}. These services
aim at aggregating data within a network in a time- and
energy-efficient manner. They are suitable when the network is
connected to a base station from which queries on sensor measurements
are issued. In TAG or TinyDB, for example, queries are entered by
means of an SQL-like syntax which tasks the network to send raw data
or aggregates at regular time intervals. These services make possible
to compute ``within the network'' common operators like
\emph{average}, \emph{min}, \emph{max}, or \emph{count}, thereby
greatly decreasing the amount of data to be transmitted. Services
typically rely on synchronized routing trees along which data is
processed and aggregated along the way from the leaves to the root
\cite{madden2002tta,madden2005taq}.

Recently, we have shown that a data aggregation service can be used to
represent sensor measurements in a different space
\cite{LeBorgne2007Unsupervised}. We suggested that the space defined
by the principal component basis, which makes data samples
uncorrelated, is of particular interest for sensor networks. This
basis is returned by the Principal Component Analysis (PCA)
\cite{hyvarinen2001ica}, a well-known technique in multivariate data
analysis. The design of an aggregation scheme which distributes the
computation of the principal component scores (i.e., the transformed
data in the PCA space) has three major benefits. First, the PCA provides
varying levels of compression accuracies, ranging from constant
approximations to full recovery of original data. Second, simple
adaptive protocols can leverage this flexibility by trading network
resources for data accuracy. Third, principal component scores
contain sufficient information for a variety of WSN applications like
approximate monitoring \cite{Li2006Smart}, feature prediction
\cite{Bontempi2005Adaptive,duarte2004vcd} and event detection
\cite{huang2006npa,lakhina2004network}.

The approach we proposed in \cite{LeBorgne2007Unsupervised}
exclusively addresses the distribution of the computation of the
principal component scores and requires the component basis to be
computed beforehand in a centralized manner. This limits the
applicability of the PCA to small networks, as the centralized computation
of the principal component basis does not scale with the network size.

The main contribution of this article is to provide a distributed
implementation of the principal component basis computation. The
algorithm is based on the Power Iteration Method~\cite{bai2000tsa,golub1996mc}, an iterative
technique for computing the principal component basis, which can be
implemented in a distributed manner by means of an aggregation
service. In particular we show that this algorithm can properly compute
the principal component basis under the mild hypothesis that the
sensor measurements collected by distant sensors are not significantly
correlated. We also extend the previous work of \cite{LeBorgne2007Unsupervised} by an in-depth discussion of
the network tradeoffs and scalability issues.

The article is structured as follows. Section \ref{section2} reviews
previous work in the domain by describing the principles of data
aggregation services in WSN and detailing how principal component
scores can be computed by an aggregation service. This section also
provides a brief overview of PCA together with potential WSN
applications and related tradeoffs. Section
\ref{section3} presents the main issues in the distribution of 
the PC basis  and proposes an implementation based on
a data aggregation service. Experimental results illustrating the
tradeoffs between accuracy and communication costs are discussed in
Section \ref{section4}. Additional related work and future research
tracks are addressed in Section \ref{section5}.

\section{Principal component aggregation in wireless sensor networks}
\label{section2}

Let us consider a static network of size $p$ whose task is to collect
at regular time instants sensor measurements or aggregates thereof,
and to route them to a destination node. This scheme is standard for
plenty of real-time WSN applications like surveillance, actuator
control or event detection. The destination node is commonly referred
to as the \emph{sink} or the \emph{base station} and is often assumed to
benefit from higher resources (e.g., a desktop computer). Let $t \in
\mathbb{N}$ refer to a discretized time domain and  $x_i[t]$ be
the measurement collected by sensor $i$, $1 \le i \le p$, at time
$t$. At each time instant $t$, the $p$ resulting measurements can be
seen as components of a vector $\B{x}[t] \in \R^p$. The sampling period is also
referred to as \emph{epoch}.

\begin{figure}[t!]
\centering
      \includegraphics[width=12cm]{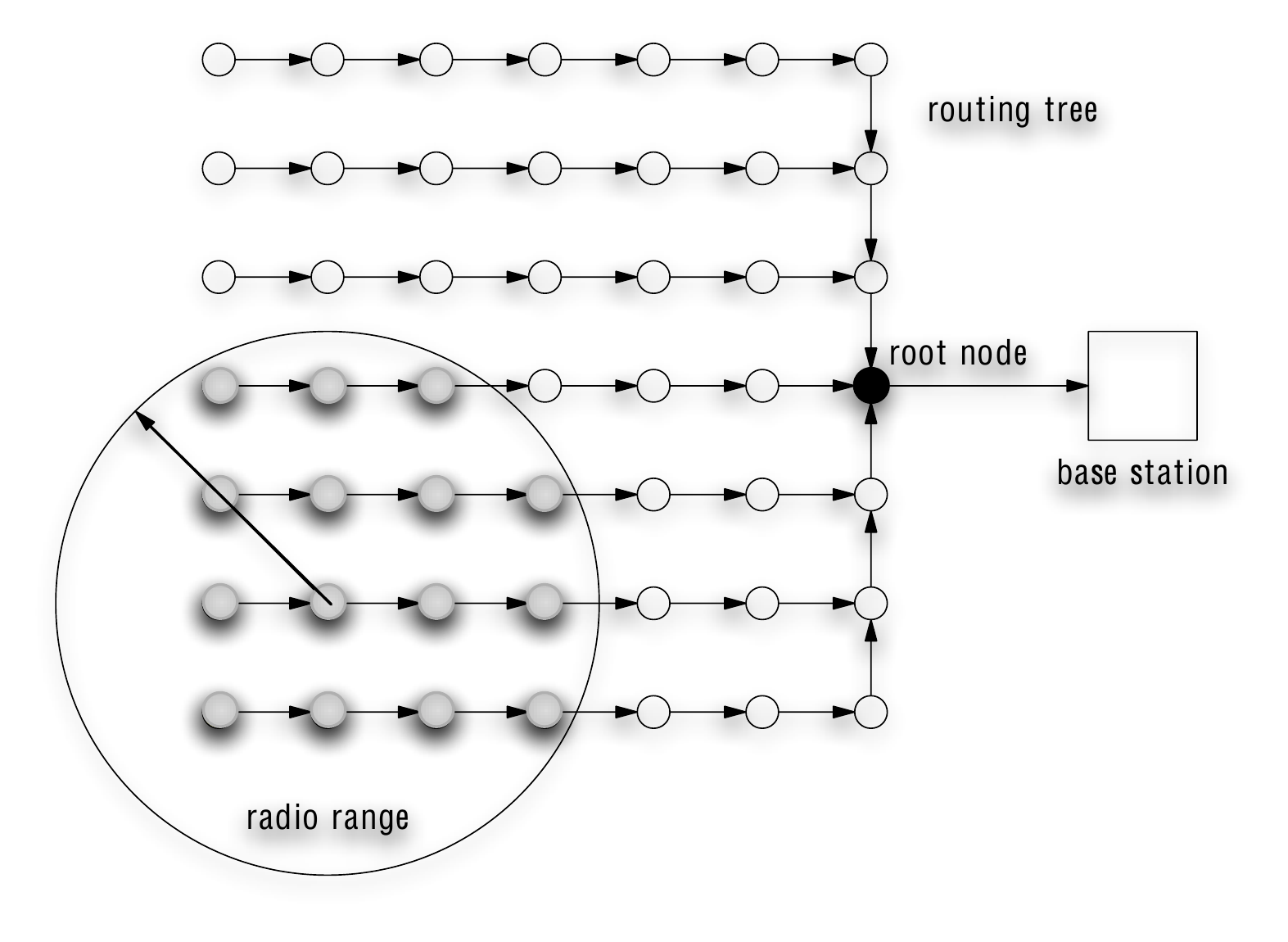}
    \caption{A multi-hop network where all the sensor nodes are
    connected to the base station by means of a routing tree. Note
    that as the radio range usually allows a node to communicate with
    several others, different routing trees can be obtained. The
    black node represents the root node.}
    \label{wsn-fig}
 \end{figure}

Since the communication range of a single node is limited, sensors
which are not in communication range of the sink have their
measurements relayed by intermediate sensors. A standard approach
consists in setting up a multi-hop network by means of a routing tree
whose root is connected to the sink (Figure \ref{wsn-fig}). Different metrics such as hop
count, latency, energy consumption and network load may be taken into
account during the design of the routing tree
\cite{akkaya2005srp}.

\subsection{Data aggregation}

\subsubsection{Aggregation service}

An aggregation service allows to aggregate data within a network in a
time- and energy-efficient manner. A well-known example of aggregation
service is the TAG system, developed at the University of Berkeley,
California \cite{madden2002tta,madden2005taq}. TAG stands for Tiny
AGgregation and is an aggregation service for sensor networks which
has been implemented in TinyOS, an operating system with a low memory
footprint specifically designed for wireless sensors \cite{tinyos}. In
TAG, an epoch is divided into time slots so that sensors' activities
are synchronized according to their depth in the routing tree. Any
algorithm can be relied on to create the routing tree, as long as it
allows data to flow in both directions of the tree and does not send
duplicates \cite{madden2002tta}. 

\begin{figure}[t!]
\centering
      \includegraphics[width=12cm]{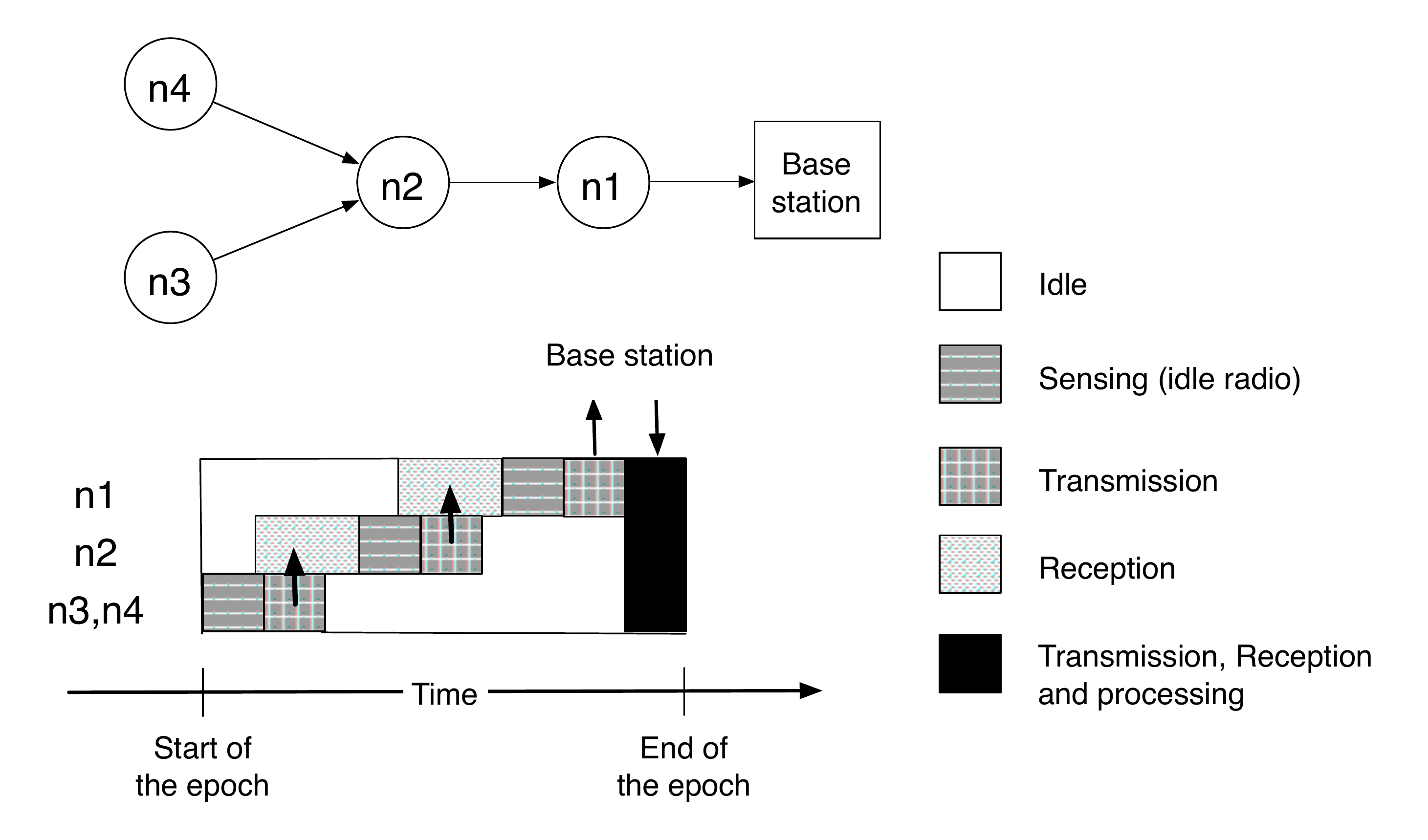}
\caption{Activities carried out by sensors in the routing tree \emph{(adapted from [13])}. Transmissions are synchronized for optimizing energy savings. The last stage involves all sensors and allows unsynchronized operations (for sensor discovery e.g.).}
\label{diag_time}
\end{figure}

The TAG service focuses on low-rate data
collection tasks which permits loose synchronization of the sensor
nodes. The overhead implied by the synchronization is therefore assumed
to be low.
The goal of synchronization is to minimize the amount of time spent by
sensors in powering their different components and to maximize the
time spent in the idle mode, in which all electronic components are
off except the clock. Since the energy consumption is several orders of
magnitude lower in the idle mode than when the CPU or the radio is
active,  synchronization significantly extends the
wireless sensors' lifetime. An illustration of the sensors' activities
during an epoch is given in Figure \ref{diag_time} for a network of
four nodes with a routing tree of depth three. Note that the synchronization is maintained at the transport layer  of the network stack, and does not require precise synchronization constraints as in TDMA. Rather, nodes are synchronized by timing information included in data packets, and a CSMA-like MAC protocol with a random backoff scheme is used at the link layer to resolve multiple access collisions. 

\subsubsection{Aggregation primitives}
\label{aggregprim}

Once the routing tree is set up and the nodes synchronized,
data can be aggregated from the leaves to the
root. Each node adds its contribution to a \emph{partial state record}
$X$ which is propagated along the routing tree. Partial state records
are merged when two (or more) of them arrive at the same node.  When
the partial state record is eventually delivered by the root node to
the base station, the desired result is returned by means of an evaluator function. An
aggregation service requires then the definition of three primitives
\cite{madden2002tta,madden2005taq}:
\begin{itemize}
\item an initializer $init$ which transforms a sensor measurement into a partial state record,
\item an aggregation operator $f$ which merges partial state records, and
\item an evaluator $e$ which returns, on the basis of the root partial
state record, the result required by the application.
\end{itemize}
Note that when partial state records are scalars or vectors, the three operators defined above may be seen as functions. Partial state records may however be any data structure which, following the notations of \cite{madden2002tta}, are represented using the symbols $ \langle . \rangle $. 

\begin{figure}
    \begin{center}
      \includegraphics[width=12cm]{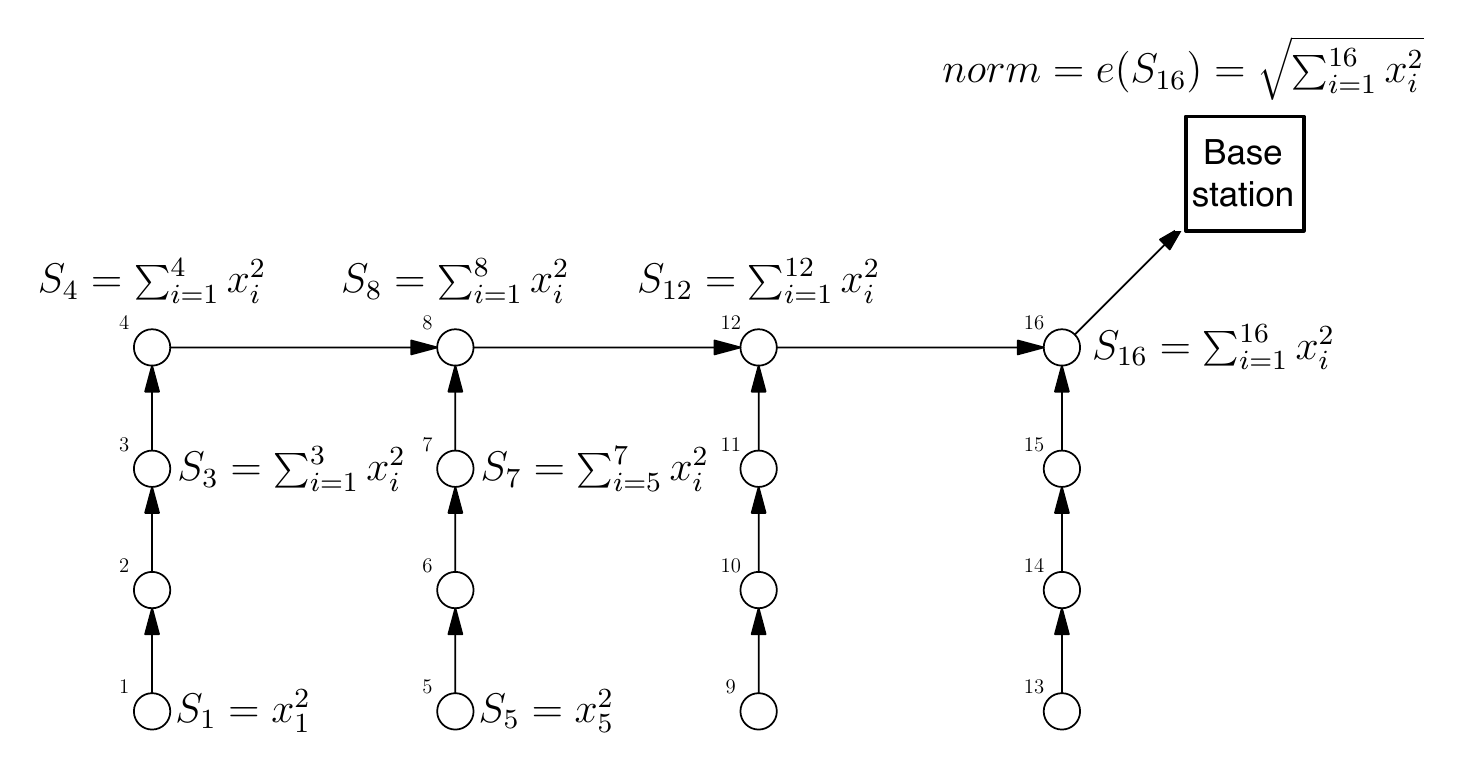}
    \end{center}
    \caption{Aggregation service at work for computing the norm of the vector measured.}
    \label{aggreg-fig}
  \end{figure}

We illustrate the aggregation principle by the following
example. Suppose that we are interested in computing the Euclidean
norm of the vector containing the WSN measurements at a given
epoch. Rather than by directly sending all the $p$ measurements to the
base station for  computation, an aggregation service (Figure
\ref{aggreg-fig}) can obtain the same result in an online manner once
the following primitives are implemented~:
  \begin{eqnarray*}
    init(x) & = & \langle x^2 \rangle \\
    f(\langle X \rangle,\langle Y \rangle) & = & \langle X+Y \rangle \\
    e(\langle X \rangle) & = & \sqrt{X}
  \end{eqnarray*}
  In this example the partial state records $X$ and $Y$ are scalars of the form
  $$
  X=\sum_{i \in I} x_i^2
  $$

\noindent where $I \subset \{1,\ldots,p\}$.

It is worth noting that in-network aggregation does not only reduce the
amount of packets transmitted but also increases the network
lifetime. The energy consumption of the processing unit of a sensor
node is at least one order of magnitude lower than the one of the
wireless radio \cite{ilyas2004hsn,polastre2005teu,zhao2004wsn}. According to the analysis \cite{raghunathan2002eaw} of
the main wireless sensor platforms available today on the market,
sending one bit of data is equivalent to $2000$ CPU cycles in terms of
energy.  For a packet length of 30 bytes (average TinyOS packet
length), this corresponds to $480000$ CPU cycles. It follows that, with
respect to transmission costs, the energy consumption of simple
aggregation operators is negligible. There
is therefore a large variety of in-network processing algorithms that
can lead to energy savings by reducing the number of packet
transmissions. 

\subsubsection{Communication costs}
\label{comcosts}

This section details the communication
costs caused by three types of network operation. The first is the \emph{default}
data collection operation in which all measurement are routed to the sink
without any aggregation. This is referred to as the D operation. The
second is the \emph{aggregation} operation, referred to as A
operation, which consists in tasking the network to retrieve an
aggregate by means of the aggregation service. Finally, we denote by
F the \emph{feedback} operation which consists in flooding the
aggregate obtained at the sink back to the whole set of sensors.

Let $q$ be the size of a partial state record, $C_i$ be the number of
direct children of node $i$ in the routing tree, $RT_i$ be the size of
the subtree whose root is the $i$-th node and $i_C^*= \argmax_i C_i$
the node whose number of children is the highest.

The following analysis compares the orders of magnitude of the
communication costs caused by the D, A and F operations,
respectively.  For this reason we consider the number of packets
processed by each node in an ideal case where overhearing, collisions
or retransmissions are ignored.

\paragraph{D operation}

Without aggregation, all the measurements are routed to the base
station by means of the routing tree. As mentioned before, the network
load at the sensor nodes, i.e., the sum of received and
transmitted packets, is ill-balanced. The
load is the lowest at by leaf nodes, which only send one packet per epoch,
while the load is the highest at the root node which processes
$2p-1$ packets ($p-1$ receptions and $p$ transmissions) per epoch. The load
at a generic $i$-th sensor node depends on the routing tree, and
amounts to $2RT_{i}-1$ packets per epoch.

\paragraph{A operation}

During the aggregation, the $i$-th node sends $q$ packets and
receives a number of packets which depends on its number $C_i$ of
children. The total number of packets processed is therefore
$q(C_i+1)$ per epoch. The load is the lowest at leaf nodes,
which only have $q$ packets to send, while the load is the highest
at the node whose number of children is the highest.

\paragraph{F operation}

The feedback operation consists in propagating the aggregated value
back from the root down to the all leaves of the tree. This operation
can be used, for instance, to get all sensor nodes acquainted with the
overall norm of their measurements or with the approximation evaluated
at the sink. The feedback of a packet containing the result of the evaluation generates a network load of two packets for all non-leaf nodes (one reception and one transmission for forwarding the packet to the children) and of one packet for the leaves (one reception only) .

\subsection{Principal component analysis}
\label{formulation}
This section describes the Principal Component Analysis (PCA), a
well-known dimensionality reduction technique in statistical data analysis, data compression,
and image processing \cite{jolliffe2002pca,Miranda2007fk}. The technique is based on a basis change which reduces the number of coordinates used to represent the data while maximizing the variance retained. Its use is particularly adapted to correlated data where the dimensionality of the data can be strongly reduced while retaining most of the measurement variations.\\
Given a set of $N$ centered multivariate measurements $\mathbf{x}[t] \in \mathbb{R}^p$ used as training samples, $1 \le t \le N$, the PCA basis is obtained by minimizing the following optimization function
\begin{equation}
\begin{array}{lll}
J_q(\mathbf{x}[t] ,\mathbf{w_k})&=&\frac{1}{N} \sum_{t=1}^N ||\mathbf{x}[t]-\sum_{k=1}^q  \mathbf{w_k} \mathbf{w_k^T}\mathbf{x}[t] ||^2 \\
&=&\frac{1}{N} \sum_{t=1}^N  ||\mathbf{x}[t]-\mathbf{\hat{x}[t]}||^2
\end{array}
\label{formulaPCA}
\end{equation}
where the set of $q \le p$ vectors $\{\mathbf{w_k}\}_{1 \le k \le q}$ of $\mathbb{R}^p$ form the PCA basis. The linear combinations $\mathbf{\hat{x}[t]}=\sum_{k=1}^q \mathbf{w_k} \mathbf{w_k}^T
\mathbf{x}[t]$ represent the projections of $\mathbf{x}[t]$ on the PCA basis, and Eq. (\ref{formulaPCA}) therefore quantifies the mean squared error the original measurements $\mathbf{x}[t]$ and their projections. Note that the measurements
are centered so that the origin of the coordinate system coincides
with the centroid of the set of measurements. This translation is
desirable to avoid a biased estimation of the basis
$\{\mathbf{w_k}\}_{1 \le k \le q}$ of $\mathbb{R}^p$ towards the
centroid of the set of measurements.

The optimization function equivalently maximizes the variance retained by the family of vectors $\{\mathbf{w_k}\}_{1 \le k \le q}$. Setting to zero the derivative of $J_q(\mathbf{x}[t],\mathbf{w_k})$ with respect to $\mathbf{w_k}$ gives
\begin{equation}
\mathbf{C} \mathbf{w_k} = \lambda_k \mathbf{w_k}
\label{PCAderiv}
\end{equation}
where $\mathbf{C}= \frac{1}{N}\sum_{t=1}^N \mathbf{x}[t] \mathbf{x}[t]^T$ is the sample covariance matrix of the sensor measurements $\mathbf{x}[t]$. The minimizer of Equation \eqref{formulaPCA}
is given by the set of the first $q$ eigenvectors $\{\mathbf{w_k}\}$ of the sample
covariance matrix $\mathbf{C}$ \cite{jolliffe2002pca}. These eigenvectors are called the \emph{principal
components} and form the PCA basis. The corresponding eigenvalues $\lambda_k$
quantify the amount of variance conserved by the eigenvectors. Indeed, left-multiplying Equation (\ref{PCAderiv}) by $\mathbf{w_k} ^T$ gives 
\begin{equation}
\mathbf{w_k} ^T\mathbf{C} \mathbf{w_k}=  \mathbf{w_k} ^T\lambda_k \mathbf{w_k}
\end{equation}
and as $\mathbf{w_k} ^T\mathbf{C} \mathbf{w_k}=\mathbf{w_k} ^T (\frac{1}{N}\sum_{t=1}^N \mathbf{x}[t] \mathbf{x}[t]^T) \mathbf{w_k}= \sum_{t=1}^N \mathbf{\hat{x}[t]}^T \mathbf{\hat{x}[t]}$ and $ \mathbf{w_k} ^T \mathbf{w_k}=1$, each eigenvalue $\lambda_k$ quantifies the variance of the projections of the measurements $\mathbf{x}[t]$ on the corresponding $k$-th eigenvector. The sum of the eigenvalues therefore equals the total variance of the original set of
observations $X$, i.e.:
\begin{equation*}
\sum_{k=1}^{p} \lambda_k = \frac{1}{N} \sum_{t=1}^{N} ||\mathbf{x}[t]||^2
\end{equation*}
It is convenient to order the vectors $\mathbf{w_k}$ by decreasing order of the eigenvalues, so that the proportion $P$ of retained variance by the first $q$ principal components can be expressed by:
\begin{equation}
P(q)=\frac{\sum_{k=1}^{q} \lambda_k}{\sum_{k=1}^{p} \lambda_k}
\label{errorPCA}
\end{equation}
Storing columnwise the set of vectors $\{\mathbf{w_k}\}_{1 \le k \le q}$ in a $W_{p \times q}$ matrix, approximations $\mathbf{\hat{x}[t]}$ to $\mathbf{x}[t]$ in $\mathbb{R}^p$ are obtained by:
\begin{equation}
\mathbf{\hat{x}[t]}=WW^T\mathbf{x}[t]=W\mathbf{z[t]}
\label{coordBack}
\end{equation}
where 

\begin{equation}
\mathbf{z[t]}=W^T\mathbf{x}[t]=\left(
    \begin{array}{l}
      \sum_{i=1}^p w_{i1} x_i \\
      ... \\
      \sum_{i=1}^p w_{iq} x_i
    \end{array}
    \right)
    = \sum_{i=1}^p \left(
    \begin{array}{l}
      w_{i1} x_i \\
      ... \\
      w_{iq} x_i
    \end{array}
    \right)
\label{coord}
\end{equation}
 denotes the column vector of coordinates of $\mathbf{\hat{x}[t]}$ in $\{\mathbf{w_k}\}_{1 \le k \le q}$, also referred to as the \emph{principal component scores}.

  \subsection{Principal component aggregation}
  \label{pcaCompr}
    
The computation of the principal component scores $\mathbf{z[t]}$ can
be performed within the network if each node $i$ is initialized with the $q$ scalar
components $\B{w}_{i1},..,\B{w}_{iq}$ of the principal component basis
\cite{LeBorgne2007Unsupervised}. As proposed in
\cite{LeBorgne2007Unsupervised}, this initialization can done during a training stage
where a set of measurements is first collected from the network and
the related covariance matrix is estimated. This set of measurements
is supposed to be large enough to provide an accurate estimate of the
covariance matrix. Once the first $q$ principal components are
computed, they are sent back to the network and stored in the
respective nodes (e.g., the $i$-th node stores only the elements
$\B{w}_{i1},..,\B{w}_{iq}$). While our proposition for distributing
the computation of the principal components will be introduced in
Section \ref{section3}, here we limit to review the aggregation
mechanism to compute the PC scores in a distributed manner.

The principal component scores can be computed by summing along the
routing tree the vectors $(\B{w}_{i1} x_i[t],..,\B{w}_{iq} x_i[t])$
available at each node. The aggregation primitives take then the
following form~:
    \begin{eqnarray*}
      init(x_i[t]) & = & \langle \B{w}_{i1} x_i[t];..;\B{w}_{iq} x_i[t] \rangle\\
      f(\langle x_1 ; .. ; x_q \rangle,\langle y_1 ; .. ; y_q \rangle) & = & \langle x_1+y_1 ; .. ; x_q+y_q \rangle\\
    \end{eqnarray*}
where partial state records are vectors of size $q$. 

\subsection{Applications}

Once the transformed data $\B{z}[t]$ reaches the base station, it can
be used for multiple purposes and applications, such as approximate
monitoring, feature extraction or event detection.

\subsubsection{Approximate monitoring}
\label{approxmon}

In approximate monitoring, the state of the process sensed by the WSN
at time $t$ is summarized  by the vector 
\begin{eqnarray*}
\hat{x}_t=e(z_1[t], ..., z_q[t])&=&(\hat{x}_1[t], ..., \hat{x}_n[t]) \nonumber\\
&=&W^Tz[t]
\end{eqnarray*}
returned by the evaluator function at the base station.

If the number of components is $p$, the exact set of measurements
collected by sensors is retrieved at the base station. Otherwise, for
a number of components less than $p$, the vector $\hat{x}_t$ provides
an optimal approximation to the real measurement vector in a mean
square sense (Equation \eqref{errorPCA}). Note that the transformation
in a principal component basis allows the design of  simple policies
to manage congestion issues, like discarding scores associated to the 
components with the lowest variance.

It is also interesting to note that a simple additional procedure can
be set up to assess the accuracy of approximations against a user
defined threshold. We know that the approximation $\hat{x}_i[t]$ for the
$i$-th sensor, $1 \le i \le p$, is given by
\eqref{coordBack} by:
\begin{equation*}
\hat{x}_{i}[t]=\sum_{k=1}^{q} z_{k}[t]*w_{ik}
\label{Checkcomp}
\end{equation*}

Since the elements $\{w_{ik}\}$ are already available at each node, by
sending back to the routing tree the $k$ principal component scores we
can enable these two additional functionalities: (i) each sensor is
able to compute the approximation retrieved at the sink, and (ii) each sensor
is able to
send a notification if the approximation error is greater than some
user defined threshold $\epsilon$. This scheme, referred to as
\emph{supervised compression} in \cite{LeBorgne2007Unsupervised},
guarantees that all data eventually obtained at the sink are within
$\pm \epsilon$ of their actual measurements.

\subsubsection{Dimensionality reduction}

An important class of WSN applications consists in inferring, on the
basis of sensor measurements, some \emph{high level} information such
as the type or number of occurred events. For instance, a
WSN monitoring the vibrations of a bridge could be used to answer queries concerning the type or number of vehicles that pass over the bridge. 
This class of problems is typically tackled by supervised learning techniques such as Naive Bayes
classifiers, decision trees, or support vector machines \cite{duda2000pc, webb1999spr}.  The PCA is in this context an effective
preprocessing technique which simplifies the learning process by reducing the dimensionality of the problem \cite{jolliffe2002pca, webb1999spr}. 

\subsubsection{Event detection}

The use of PCA for identifying at the network scale unexpected events
which are not detectable at the node level has been discussed in the literature
\cite{huang2006npa,lakhina2004network}. Detection of such events can
be achieved by focusing on the low variance components. In fact, such
components are expected to bear coordinates close to zero, as they
typically account for sensor equipment noise. In this context, the
principal component aggregation scheme can be used to deliver the value of low
variance components, and the evaluator
function takes the form of a statistical test checking if the
coordinates on one or more low variance components is different from
zero.

\subsection{Tradeoffs}
\label{tradeoffPCA}

In this section, we will use the term \emph{default} to denote the scheme that forwards all the
measurements to the sink at each epoch, and the term
\emph{PCAg} to denote the scheme based on the principal component
aggregation. Let us first consider the distribution of the network load among sensor nodes, with a particular focus on its upper bound. The upper bound, henceforth referred to as \emph{highest network load}, determines the network capacity, that is the amount of traffic that the network can handle at one time \cite{ilyas2004hsn}. The highest network load is also related to  the network lifetime in terms of
time to first failure (i.e., the time at which the first node in the network runs out of
energy), as the radio is known to be the most energy consuming task of a
wireless sensor \cite{polastre2005teu,akyildiz2002wsn}. 

In the \emph{default scheme}, the root of the routing tree is the node that sustains the highest network load as it forwards to the base station the
measurements from all other sensors. Its radio throughput therefore
causes the root node to have its energy exhausted first, which may lead the rest of the network to be disconnected from the base station.

In the \emph{PCAg scheme} (Section \ref{comcosts}), the network load is distributed in a more homogeneous manner. 
The node ${i_\mathcal{C}^{*}}= \argmax_i C_i$ whose number of children is the highest 
becomes the limiting node in terms of highest network load. The network traffic at this node
amounts to $q(C_{i_\mathcal{C}^{*}}+1)$ packets per epoch, where $q$ is the number
of components retained. A PCAg scheme therefore reduces the highest network load in a configuration where 
\begin{equation}
q(C_{i_\mathcal{C}^{*}}+1)\le 2p-1.
\label{tradeoff}
\end{equation}
We illustrate the tradeoff by considering two extreme configurations. If only the first principal component is retained, the PCAg reduces the highest network load as $C_{i_\mathcal{C}^{*}}<p$. On the contrary, if all the components are retained, i.e. $q=p$, the equation \eqref{tradeoff} simplifies into $pC_{i_\mathcal{C}^{*}} \le p-1$, and the PCAg incurs a higher network load than the default scheme. 

The main parameter of the approach is therefore the number of principal components to retain, which trades network load for sensing task accuracy. The amount of information retained by a set of $q$ principal components depends on the degree of
correlation among the data sources. Whenever a set of sensors collects
correlated measurements, a small proportion of principal components is
likely to support most of the variations observed by the network. The
number of principal components to retain can be specified by the
application, for example by means of a cross validation procedure on
the accuracy of the sensing task at the base station \cite{LeBorgne2007Unsupervised}. It can be also
be driven by network constraints or network
congestion issues. In the latter case, a policy prioritizing packets
containing the component scores of the principal components would
allow to retain a maximum amount of information under network load
constraints.

Finally, given a number of principal components, note that the highest network load may be further reduced if the algorithm that builds the routing tree uses $C_{i_\mathcal{C}^{*}}$ as a parameter. The network load indeed depends on the number of children a node has. Although we de not address this aspect, routing strategies that aim at lowering $C_{i_\mathcal{C}^{*}}$ can therefore further improve the efficiency of the PCAg scheme. 

\section{Computation of the principal components}
\label{section3}

\subsection{Outline}
  
This section addresses the initialization procedure whereby  each node $i$ can be initialized with the elements
$\B{w}_{i1},..,\B{w}_{iq}$ of the principal components $\B{w}_1,..,\B{w}_q$. In \cite{LeBorgne2007Unsupervised}, we proposed a centralized approach for performing this initialization stage. This approach is first briefly recalled in \ref{centralized}. We then extend this work by showing that a fully distributed implementation can perform this initialization process. The proposed approach relies on the power iteration method (PIM), a classic iterative technique for computing the eigenvectors of a matrix \cite{bai2000tsa}, and on a simplifying assumption on the covariance structure which allows to estimate the covariance matrix in a distributed way. Section  \ref{onlineCov} presents and discusses the distribution of the covariance matrix computation and storage. We then show in Section \ref{pim} that by means of the distributed covariance matrix  computation, the PIM can be implemented in an aggregation service in a distributed manner.

Each of these sections is concluded by a complexity analysis of the communication, computational,  and memory costs. We finally discuss in Section \ref{summaryPIM} three methods for performing the initialization stage.

  \subsection{Centralized approach}
  \label{centralized}

Let us suppose that $t$ vectors of sensor measurements $\B{x}[\tau]
\in \R^p$, $\tau \in \{1,..,t\}$ are retrieved at the
base station. Let $\B{X}[t]$ denote the $p \times t$ matrix whose
$\tau$-th column-vector is the vector $\B{x}[\tau] \in \R^p$. An
estimate of the covariance matrix ${\B{C}[t]=(c_{ij}[t])_{1 \leq i,j
\leq p}}$ is obtained at time $t$ by computing
\begin{eqnarray}
  \B{C}[t] & = &\frac{1}{t} \sum_{\tau=1}^t (\B{x}[\tau]-\mathbf{\bar{x}}[t])(\B{x}[\tau]-\mathbf{\bar{x}}[t])^T \nonumber \\
&=&\frac{1}{t} \B{X}[t]\B{X}[t]^T- \mathbf{\bar{x}}[t]\mathbf{\bar{x}}[t]^T
\label{eqcov}
\end{eqnarray}
where $\mathbf{\bar{x}}[t]=\frac{1}{t} \sum_{\tau=1}^{t} \mathbf{x}[\tau]$ is the average vector.

The covariance matrix can be updated recursively as new vectors of
observations $\B{x}[t+1]$ are made available at the base
station. From Equation \eqref{eqcov} it follows that
  \begin{eqnarray}
  c_{ij}[t] & = & \frac{1}{t} S_{ij}[t]- \frac{1}{t^2}S_i[t]S_j[t]
  \end{eqnarray}
  where
  \begin{eqnarray}
   S_i[t] & = & \sum_{\tau=1}^{t}x_i[\tau] = S_i[t-1] + x_i[t] \nonumber \\
   S_{ij}[t] & = & \sum_{\tau=1}^t x_i[\tau] x_j[\tau] = S_{ij}[t-1] + x_i[t] x_j[t]
\label{covrec}
  \end{eqnarray}
This means that the storage of the quantities  $t$, $S_{j}$ and $S_{ij}$, $1 \le i,j \le p$, 
in the base station makes possible the update of the covariance matrix estimation $\B{C}[t]$ at time $t$. 
Once the covariance matrix is estimated, the principal components can be computed 
using a standard eigendecomposition method~\cite{bai2000tsa} and its elements communicated to the sensor nodes.

\subsubsection{Scalability analysis}

\paragraph{Highest network load:} 
The centralized estimation of the covariance matrix from a set of $t$ vectors 
of observations requires $t$ operations of type D as defined in Section \ref{comcosts}. 
The root node supports the highest network load, which is therefore in $O(tp)$.  Once the eigenvector decomposition is performed at the base station, the root node then transmits the principal components to the sensor nodes, which requires $qp$ operations of type $F$.  Overall, the communication cost amounts to $O(tp+qp)$.

\paragraph{Computational and memory costs:} 
From Equation \eqref{covrec}, the computational cost related to the covariance matrix at the base station is $O(tp^2)$.
The memory cost for storing the matrix of observations and the covariances is
$O(tp+p^2)$ or $O(p^2)$ if recursive updates are relied on. As far as
the estimation of the principal components is concerned, the cost of
standard eigendecomposition algorithm is $O(p^3)$ in terms of computation
and $O(p^2)$ in terms of memory~\cite{bai2000tsa}.

  \subsection{Distributed estimation of the covariance matrix}
  \label{onlineCov}
This section presents an alternative methodology to distribute the
computation and storage of the covariance matrix within the
network. The methodology relies on the observation that sensors that
are geographically distant often present low correlation values. This hypothesis often holds in real-world situations as shown in \cite{barrenetxea2007slu,jindal2004msc}. This
leads us to make the simplifying hypothesis that covariances between
sensors which are out of reach is equal to zero. Note that
the radio range can usually be tuned on wireless sensor platforms by
increasing or reducing the power transmission level
\cite{polastre2005teu}. If available, this feature could be used to
set the radio range such that nearby
correlated sensors are within radio range.

Let $\mathcal{N}_i$ be the neighborhood of the sensor $i$, i.e., the set
of sensors which are in the communication range of the $i$-th node. An instance of neighborhood is illustrated in
Figure \ref{wsn-fig}. Our simplifying hypothesis assumes that $c_{ij}[t]=0$ for all $j
\notin \mathcal{N}_i$. This makes the computation of the
covariance matrix scalable to large networks. The covariances between
sensors $i$ and $j$ can indeed be computed recursively by Equation
\eqref{covrec} once the sensor $i$ keeps over time a record of $t$, $S_{j}$
and $S_{ij}$ for $j \in \mathcal{N}_i$. This hypothesis will be referred to in the following as the \emph{local covariance hypothesis}.

\subsubsection{Positive semi-definiteness criterion}

If the local covariance hypothesis is not verified, the distributed sparse matrix obtained by the
local covariance hypothesis may lead to a globally non positive
definite matrix (i.e., that contains negative eigenvalues). The positive semi-definiteness is however a necessary (and sufficient)
condition for a symmetric matrix to be a covariance matrix. To face this issue, the
approach adopted in this article is to discard eigenvectors whose
eigenvalues are found to be negative. This approach, according to~\cite{rousseeuw1993tnp}, is the  simplest way to
transform a non positive definite matrix in a semi positive definite
matrix. The detection of a negative eigenvalue will be detailed in Section \ref{pimortho}, and used as a stopping criterion in the last line of Algorithm \ref{algopim}.

A good review of existing methods for transforming the
approximated covariance matrix can be found in
\cite{rousseeuw1993tnp}. Our opinion is that it seems however difficult to implement these
method in a distributed manner. Note that despite its simplicity, the
method that consists in discarding negative eigenvectors was observed
by authors in~\cite{rousseeuw1993tnp} to provide good approximations
to the covariance matrix.

\subsubsection{Scalability analysis}
\label{scalecov}

\paragraph{Highest network load:} Each update requires a node $i$ to send one packet (its measurement) and to receive $|\mathcal{N}_i|$ packets (other sensors' measurements). Let $i_\mathcal{N}^{*}$ be the node that has the largest number of neighbors, i.e.,  $i_\mathcal{N}^{*}= \argmax_i |\mathcal{N}_i|$. The  highest network load is therefore $O(t|\mathcal{N}_{i_\mathcal{N}^{*}}|)$.

\paragraph{Computational and memory costs:} 
From Equation \eqref{covrec}, the updates of $S_j$ and $S_{ij}$ demand $3|\mathcal{N}_i|$
floating points additions/subtractions, $2|\mathcal{N}_i|$ floating points
multiplications and $2|\mathcal{N}_i|$ floating points divisions. The highest 
computational cost therefore scales in $O(|\mathcal{N}_{i_\mathcal{N}^{*}}|)$. The
number of variables to maintain at a sensor node is one integer and
$2|\mathcal{N}_{i}|$ floating points numbers, so the highest memory cost amounts to $O(|\mathcal{N}_{i_\mathcal{N}^{*}}|)$.

  \subsection{Distributed estimation of the principal components}
  \label{pim}
  
The distribution of eigendecomposition algorithms is a fairly known
research area~\cite{bai2000tsa}. However, the distribution sought in our problem
should comply with the specificity that each node only has access to a single dimension of the problem and needs 
a specific entry of each principal eigenvector.

We propose in the following an approach that leverages the aggregation
service to implement in a rather simple way the power iteration method
(PIM), a well-known solution to the eigenvalue
decomposition problem \cite{bai2000tsa}. The resulting technique complies with
the specific features mentioned above, and lets each node $i$  locally
identify the subset $\{w_{ik}\}$ required by the principal component
aggregation.

After describing the PIM in Sections \ref{pimalgo} and \ref{pimortho},
we detail in Sections \ref{approxPIM} and \ref{pimsynchro} its
implementation in the aggregation service. The analysis of complexity
of the method is provided in Section \ref{pimto}.

    \subsubsection{Power iteration method}
  \label{pimalgo}

Let $\B{C}$ be a covariance matrix. The power iteration method starts
with a random vector $\B{v}_0$. At the $t$-th iteration, the vector
$\B{v}_{t+1}$ is obtained by multiplying $\B{C}$ by $\B{v}_t$, and by
normalizing it. It can be easily shown~\cite{bai2000tsa} that $\B{v}_t$
converges to the principal eigenvector $\B{w}_1$ of $\B{C}$, under the
mild condition that $\B{v}_0$ is not strictly orthogonal to the
principal eigenvector. The convergence rate is exponential, its base
being the squared ratio of the two principal eigenvalues. The
convergence criteria can be defined either as a minimum variation
$\delta$ for $\B{v}_t$, or as a highest number of iterations
$t_{\mathrm{max}}$ \cite{bai2000tsa,golub1996mc}. Algorithm
\ref{classicalPIM} outlines the different steps.

  \begin{algorithm} 
  \caption{Standard power iteration algorithm}
  \label{classicalPIM}
  \begin{algorithmic}
   \STATE $\B{v}_0 \leftarrow \mathrm{arbitrary~initialization}$
   \STATE $t \leftarrow 0$
   \REPEAT
    \STATE $\B{v}_{t} \leftarrow \B{C}\B{v}_t$
    \STATE $\B{v}_{t+1} \leftarrow \frac{\B{v}_{t}}{\norm{\B{v}_{t}}}$
    \STATE $t \leftarrow t+1$
   \UNTIL{ $t > t_{max} ~\mathrm{and/or} ~\norm{\B{v}_{t+1} - \B{v}_{t}} \leq \delta $ }
   \STATE return $\B{v}_t$
  \end{algorithmic}
  \end{algorithm}
  
Note that as the method converges to the principal eigenvector
$\B{w}_1$, the normalizing factor $\norm{\B{v}_t}$ converges to the
associated eigenvalue $\lambda_1$, as by definition~:

\begin{equation}
\B{C}\B{w}_1=\lambda_1\B{w}_1
\label{eig}
\end{equation}

  In practical settings, the power
method quickly converges to a linear combination of eigenvectors whose
eigenvalues are close, or to the eigenvector whose eigenvalue is the
highest if eigenvalues are well separated. As our purpose
here is to find the subspace that contains the signal, eigenvectors
with close eigenvalues can be thought of as generating a subspace
where similar amount of information are retained along any vector of
the subspace. The convergence to a linear combination of eigenvectors
with close eigenvalues is therefore deemed acceptable as far as
principal component aggregation is concerned.

\subsubsection{Computation of subsequent eigenvectors}
\label{pimortho}

The standard way to employ PIM in order to find the other eigenvectors (up to the number $q$)
is the deflation method which consists in applying the PIM to the
covariance matrix from which we have removed the contribution of the
$k$ principal eigenvector already computed~\cite{bai2000tsa}.
We obtain
 \begin{eqnarray*}
\B{v}_{t+1}&=& (\B{C}- \sum_{l=1}^{k}\B{w}_l\lambda_l\B{w}_l^T)\B{v}_t\\
 &=&  \B{C}\B{v}_t- \sum_{l=1}^{k}\B{w}_l\lambda_l\B{w}_l^T\B{v}_t
  \end{eqnarray*}
which boils down to orthogonalize $\B{v}_t$ with respect to all previously
identified eigenvectors $\{\B{w}_l\}_{1\le l < k+1}$. 

A verification step must be added to the algorithm in order to detect
negative eigenvalues.
This can be achieved by comparing the sign of elements of $\B{v}_t$ and $\B{v}_{t+1}$ after
convergence is assumed. From Equation \eqref{eig}, a negative eigenvalue implies that elements of $\B{v}_t$ and $\B{v}_{t+1}$ have pairwise different signs. The criterion we define for determining the sign of an eigenvalue is 
$$
sign(\sum_{i=1}^p sign(v_t[i]v_{t+1}[i])
$$
as taking the average sign of the sum of the signs over all pairwise products of $\B{v}_t$ and $\B{v}_{t+1}$ makes a more robust estimate of the sign of the eigenvalue in case of numerical errors. We rely on this additional criterion for stopping the computation of the principal components. The resulting
process is given by Algorithm \ref{modifPIMq}.

   \begin{algorithm} 
    \caption{Modified power iteration algorithm for $q$ eigenvectors}
    \label{modifPIMq}
    \begin{algorithmic}
      \STATE $k \leftarrow 0$
      \REPEAT
	\STATE $k \leftarrow k+1$
      \STATE $t \leftarrow 0$
	\STATE $\B{v}_0 \leftarrow \mathrm{arbitrary~initialization}$
	\REPEAT
    \STATE $\B{v}_{t} \leftarrow \B{C}\B{v}_t$
	  \STATE $\B{v}_{t} \leftarrow \B{v}_{t} - \sum_{l=1}^{k-1}\langle \B{v}_{t} , \B{w}_l \rangle \B{w}_l$
    \STATE $\B{v}_{t+1} \leftarrow \frac{\B{v}_{t}}{\norm{\B{v}_{t}}}$

	  \STATE $t \leftarrow t+1$
	\UNTIL { $t > t_{\mathrm{max}}$ or $\norm{\B{v}_{t+1} - \B{v}_{t}} \leq \delta$ }
	\STATE $\lambda_k \leftarrow \pm \norm{\B{v}_{t}}$
	\STATE $\B{w}_k \leftarrow \B{v}_{t+1}$
	\UNTIL {$k=q~\mathrm{or}~\lambda_k < 0$}
    \end{algorithmic}
    \label{algopim}
  \end{algorithm}

    \subsubsection{Implementation in the aggregation service}
    \label{approxPIM}
  
    A fully distributed implementation of the algorithm is possible if the covariance matrix has been estimated as proposed in \ref{onlineCov}. The computation of one iteration therefore goes in two stages~:
    
    \paragraph{Computation of $\B{C}\B{v}$}
    
    $$\forall i \in \{1,..,p\}~,~\left(\B{C}\B{v}\right)[i] = \sum_{j=1}^p c_{ij}v[j]$$
    
\noindent    As we made the assumption that:
    
    $$
    \forall i \in \{1,..,p\}~,~\forall j \not\in \mathcal{N}_i~,~c_{ij}=0
    $$
    
\noindent    The sum can be simplified~:
    $$\forall i \in \{1,..,p\}~,~\left(\B{C}\B{v}\right)[i] = \sum_{j \in \mathcal{N}_i} c_{ij}v[j]$$
    
\noindent    Recall that each node $i$ has available the set of covariances $(c_{ij})$, $j \in \mathcal{N}_i$.  Therefore at each step $t$, node $i$ only has to broadcast $v_t[i]$ and to receive the $v_t[j]$, $j \in \mathcal{N}_i$ from its neighbors for updating locally its $\sum_{j \in \mathcal{N}_i} c_{ij}v[j]~=~\left(\B{C}\B{v}\right)[i]$. The power iteration method can thus be implemented in a fully distributed manner.
    
    \paragraph{Normalization and orthogonalization}
The norm can be computed using the primitives detailed in Section \ref{aggregprim}, while the orthogonalization is obtained by means of these primitives:  
  
  \begin{eqnarray*}
    init(v_{t}[i]) & = & \langle (v_{t}[i]\B{w_l}[i])_{1 \le l \le k-1} \rangle \\
    f(\langle X \rangle,\langle Y \rangle) & = & \langle X+Y \rangle \\
    e(\langle X \rangle) & = & X.
  \end{eqnarray*}

The resulting $k-1$ scalar products $\{\langle \B{w}_l,\B{v_{t}}\rangle\}_{1 \le l \le k-1}$ are then communicated to the sensors by means of an F operation, so that each sensor $i$ can locally orthogonalize  $\B{v}_{t+1}[i] \leftarrow \B{v}_{t}[i]  - \sum_{l=1}^{k-1}\langle \B{v}_{t}  , \B{w}_l \rangle \B{w}_l[i]$. 

    \subsubsection{Synchronization}
\label{pimsynchro}  

      The nodes have to be synchronized so that they all work on the same copy of $\B{v}$ at the same time. The scheduling policy is presented in Algorithm \ref{distrPIMsync}. The implementation is straightforward since the aggregation service is synchronized.
      
      \begin{algorithm}
        \caption{Network synchronization scheme for the distributed power iteration algorithm}
        \label{distrPIMsync}
        \begin{algorithmic}
	  \STATE $k \leftarrow 0$
	  \REPEAT
	    \STATE $\forall i$ node $i$ is initialized with $C[i,i]$
	    \REPEAT
	      \STATE The nodes get their neighbor's value $(\B{v_t}[j])_{j \in \mathcal{N}_i}$
	      \STATE The computation of $C\B{v_t}$ is performed in parallel
	      \STATE $\norm{\B{v_{t}}}$ and $\{\langle \B{w}_l,\B{v_{t}}\rangle\}_{1 \le l \le k-1}$ are computed by the aggregation service
	      \STATE $\norm{\B{v_{t}}}$ and $\{\langle \B{w}_l,\B{v_{t}}\rangle\}_{1 \le l \le k-1}$ are fed back in the network
	      \STATE For all $i$, $\B{v_{t+1}}[i] \leftarrow \frac{(C\B{v_t})[i]- \sum_{l=1}^{k-1}\langle \B{v}_{t}  , \B{w}_l \rangle \B{w}_l[i]}{\norm{\B{v_t}}}$ (in parallel on all nodes)
	    \UNTIL {convergence is obtained}
	  \STATE $\B{w}_k \leftarrow \B{v}$
	  \UNTIL {$k=q$ or $\lambda_k \le 0$}
        \end{algorithmic}
      \end{algorithm}

\subsubsection{Scalability analysis}
\label{pimto}

\paragraph{Highest network load:} At each iteration, the computation of the product $\B{C}\B{v_t}$ requires a node $i$ to send one packet and to receive $|\mathcal{N}_i|$ packets. The highest network load for this operation is therefore $O(|\mathcal{N}_{i_\mathcal{N}^{*}}|)$. The normalization step implies one A and one F operations (as defined in \ref{comcosts}), and the orthogonalization step implies $k-1$ operations of type A and F respectively, where $k$ is the index of the principal component computed. The highest network load for the computation of the $q$ first principal components amounts to $O(q|\mathcal{N}_{i_\mathcal{N}^{*}}|+q^2|\mathcal{C}_{i_\mathcal{C}^{*}}|)$. 

The highest network load related to the power iteration method is therefore quadratic in the number of principal components computed. Depending on the constraints of the network, it may be more interesting to retrieve the approximated covariance matrix from the network at the base station for a centralized computation of the eigenvectors. The tradeoff depends here on the communication and computational resources available in the network, and will be discussed more in detail in Section \ref{summaryscheme}.

 \paragraph{Computational and memory costs} For a node $i$, the cost of the computation of $\B{C}\B{v_t}$ is $O(|\mathcal{N}_i|)$. The cost of the normalization step is $O(|\mathcal{C}_i|)$ for node $i$, and the orthogonalization step is  $O(k|\mathcal{C}_i|)$. The overall highest computational cost therefore amounts to $O(q(|\mathcal{N}_{i_\mathcal{N}^{*}}|+|\mathcal{C}_{i_\mathcal{C}^{*}}|)$. Regarding memory costs, each node $i$ needs to maintain variables for storing its local $w_{ik}$  and its neighbors parameters $w_{jk}$, $j \in \mathcal{N}_i$. The complexity of the highest memory cost is therefore $O(q+|\mathcal{N}_{i_\mathcal{N}^{*}}|)$.

\subsection{Summary}
\label{summaryPIM}

\label{summaryscheme}

The benefits of the different approaches for computing the covariance matrix and its principal components depend on the communication, computational and memory constraints available in the network nodes and at the base station. A summary of the complexities of these approaches is given in Table \ref{complexity}.  

\begin{table}[ht]
\begin{center}
\begin{tabular}{|l|l|l|l|}
  \hline
Operation & Communication & Computation & Memory  \\
  \hline
  \hline
\raggedleft{Covariance} & & & \\
  \hline
\raggedright{Centralized} & $O(pT)$ & $O(p^2T)$  &  $O(p^2)$ \\
\raggedright{Distributed} & $O(|\mathcal{N}_{i_\mathcal{N}^{*}}|T)$& $O(|\mathcal{N}_{i_\mathcal{N}^{*}}|T)$&  $O(|\mathcal{N}_{i_\mathcal{N}^{*}}|)$ \\
  \hline
\hline
Eigenvectors & & & \\
  \hline
Centralized & $O(qp)$&$O(p^3)$ &$O(p^2) $\\
Distributed &$O(q^2|\mathcal{N}_{i_\mathcal{N}^{*}}|)$& $O(q(|\mathcal{N}_{i_\mathcal{N}^{*}}|+|\mathcal{C}_{i_\mathcal{C}^{*}}|))$& $O(q+|\mathcal{N}_{i_\mathcal{N}^{*}}|)$\\
   \hline
 \end{tabular}
\end{center}
\caption{Scalability of the centralized, hybrid and distributed schemes.}
 \label{complexity}
\end{table}

\section{Experimental results}
\label{section4}

This section illustrates by means of a compression task the ability of the proposed approach to properly identify the principal component subspace, and discusses the tradeoffs involved between the compression accuracy and the communication costs. The experimental study is based on a set of real world temperature measurements involving a network of 52 sensors (Section \ref{dataset}), and on a network simulation used to vary the structure of the routing trees to study the impact of the network topology on the network load (Section \ref{netsim}). Results on the ability of a few components to retain most of the information are first illustrated in Section \ref{respcaggacc}, followed in Section \ref{respcaggnet} by an analysis of the network loads as the number of retained components increases. The ability of the distributed PCA to properly identify the principal components is then studied in Sections \ref{sup} and \ref{resdpca}, which focus on the covariance matrix approximation and eigenvector extraction stages, respectively. Section \ref{resdisc} concludes by a discussion on the main results.

\subsection{Dataset description}
\label{dataset}

Experiments were carried out using a set of five days of temperature readings obtained from a 54 Mica2Dot sensor deployment at the Intel research laboratory at Berkeley \cite{intellab}. The sensors $5$ and $15$ were removed as they did not provide any measurement. The readings were originally sampled every thirty-one seconds. A preprocessing stage where data was discretized in thirty second intervals was applied to the dataset. After preprocessing, the dataset contained a trace of 14400 readings from 52 different sensors. The code associated to the preprocessing and the network simulation was developed in R, an open source statistical language, and is available from the authors' web site \cite{labowsn}.

\begin{figure}[!ht]
\begin{minipage}{.45\textwidth}
\centering
    \includegraphics[width=6cm]{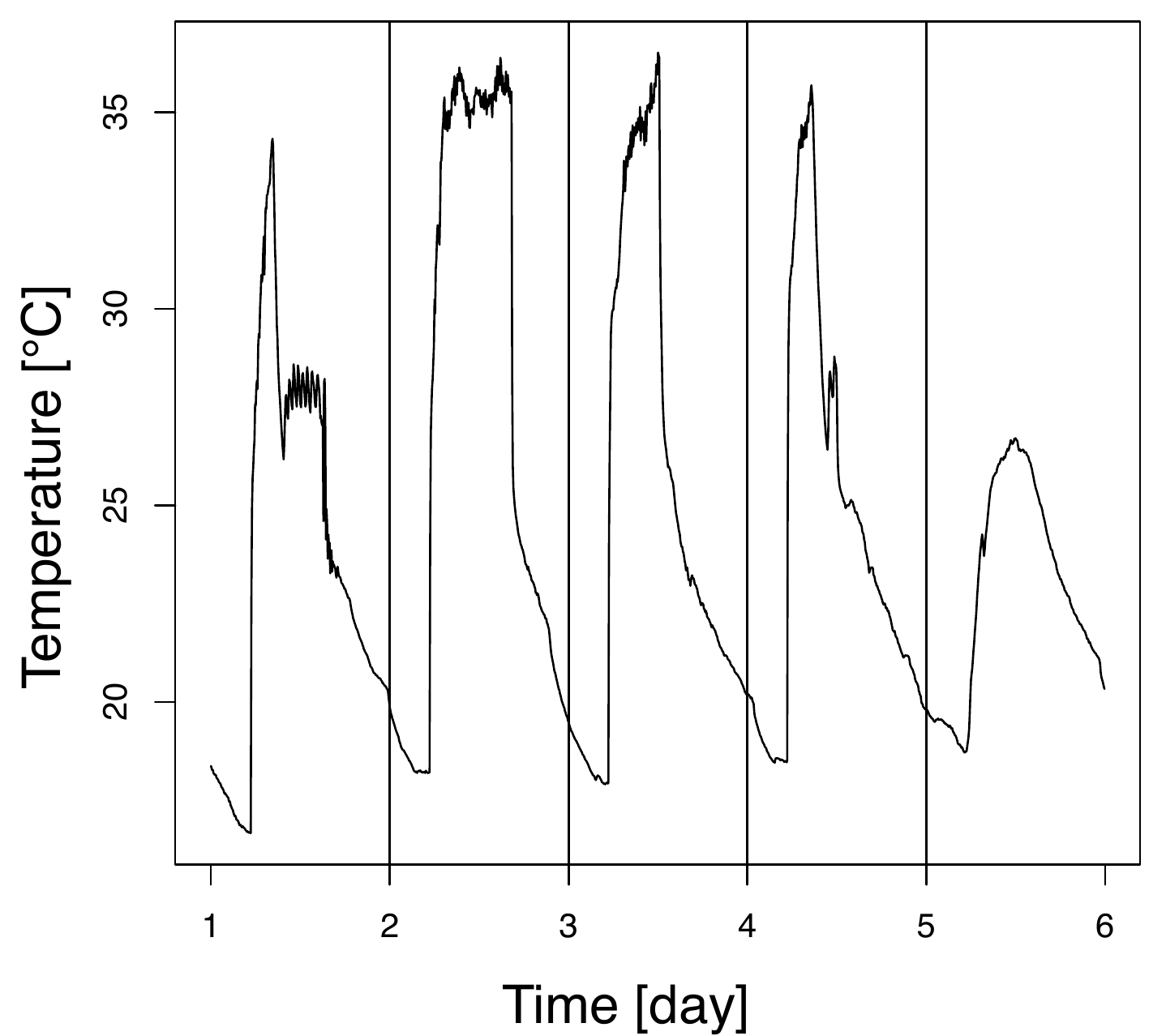}
\caption{Temperature measurements collected by sensor 21 over a five day period.}
\label{figProfileIntel}
\end{minipage}
\hfill
\begin{minipage}{.45\textwidth}
\centering
    \includegraphics[width=6cm]{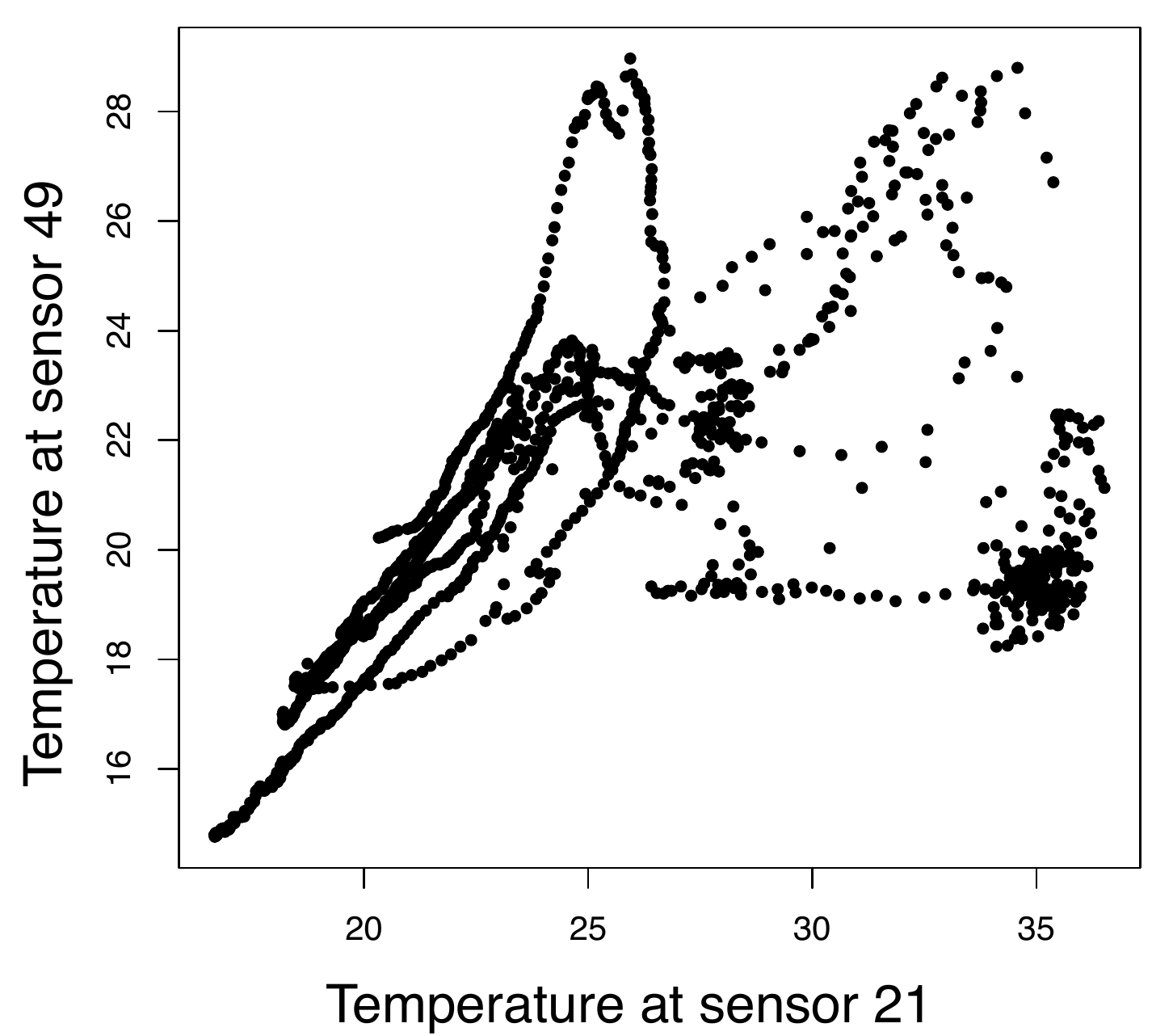}
\caption{Examples of the dependencies between the measurements of sensor 21 and sensor 49.}
\label{figSensCorIntel}
\end{minipage}
\end{figure}

Examples of temperature profiles and dependencies between measurements are reported in Fig. \ref{figProfileIntel} and \ref{figSensCorIntel}, respectively. The sensors 21 and 49 were the least correlated ones over that time period, with a correlation coefficient of $0.59$. They were situated on opposite sides of the laboratory. Temperature over the whole set of data ranged from about $15^\circ$C to $35^\circ$C.

\subsection{Network simulation}
\label{netsim}

\begin{figure}[ht]
    \centering
      \includegraphics[width=10cm]{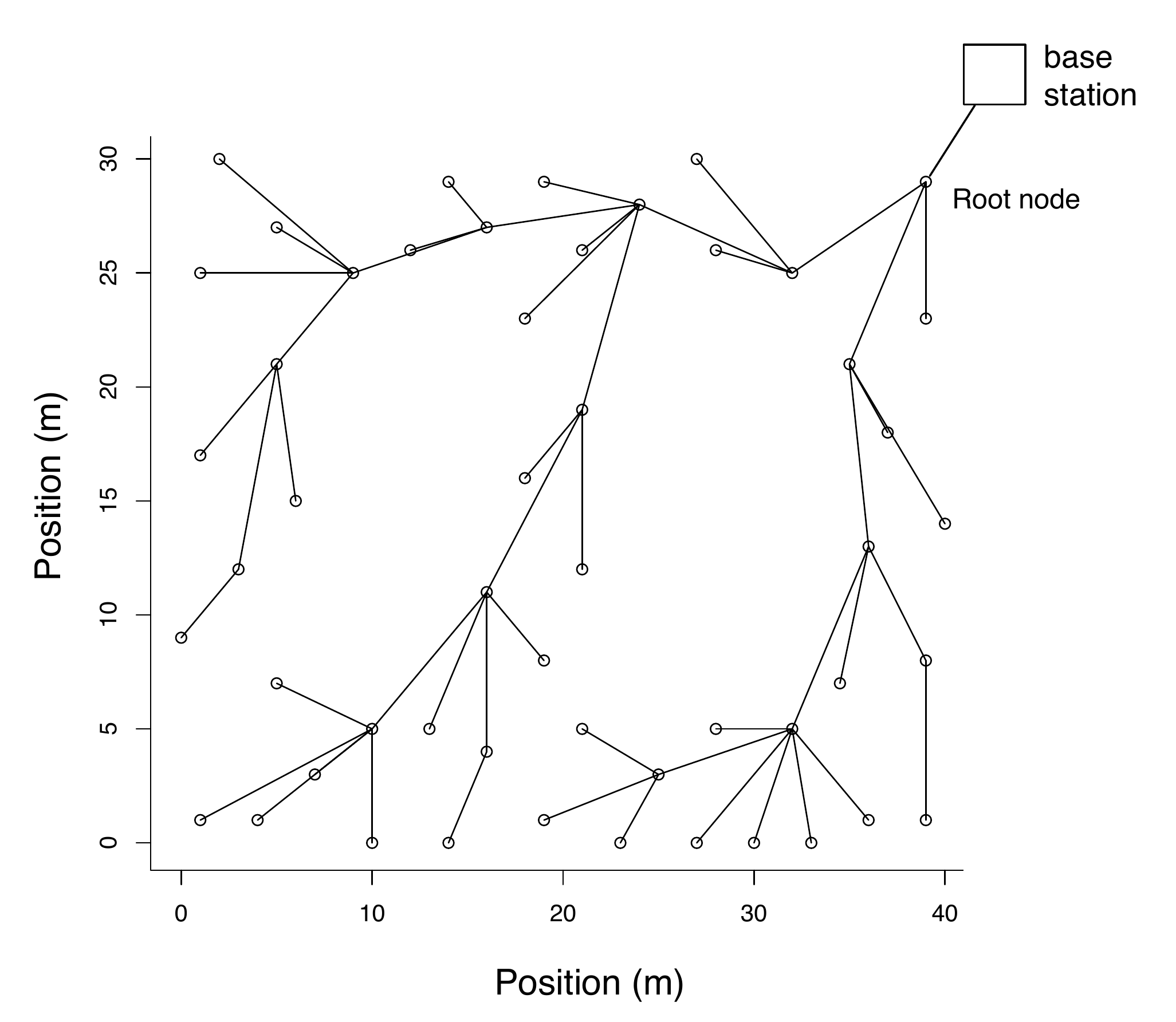}
\caption{Map of sensors and the routing obtained for a connectivity of 10 meters. The root node connecting the sink is the top right sensor. The routing tree has a depth of seven.}
\label{routingTree}
\end{figure}

The positions of the sensors are provided in \cite{intellab}, and the distribution of the sensors in the laboratory can be seen in Fig. \ref{routingTree}. We analyzed the communication costs in different routing trees which were generated in the following way. The root node was always assumed to be the top right sensor node in Fig.  \ref{routingTree} (node $16$ in \cite{intellab}). The routing trees were generated on the basis of the sensor positions and the radio range was varied from $6$ meters (minimum threshold such that all sensor could find a parent) to $50$ meters (all sensors in radio range of the root node). Starting from the root node, sensors were assigned to their parent in the routing tree using a shortest path metric, until all sensors were connected. An illustration of the routing tree obtained for a maximum communication range of $10\rm{m}$ is reported in Fig.  \ref{routingTree}.

\subsection{Principal component aggregation}
\label{respcaggacc}

We study in this section the amount of variance that the PCA can retain in the measurements for different training periods and number of components. We rely on a 10-fold cross validation technique to simulate the fact that the measurements used to compute the principal components are not used for assessing the accuracy of the approach. More precisely,  the dataset is split in ten consecutive blocks (1440 observations -- i.e., half a day of measurements). Each of the ten blocks is used in turn as a \emph{training} set to compute the covariance matrix and its eigenvectors, while the remaining observations, referred to as \emph{test} set, are used to estimate the percentage of retained variance. This provides ten estimates of the percentage of retained variance on measurements not used for computing the eigenvectors, which we average to obtain an overall estimate of the method.

\begin{figure}[!ht]
    \centering
      \includegraphics[width=10cm]{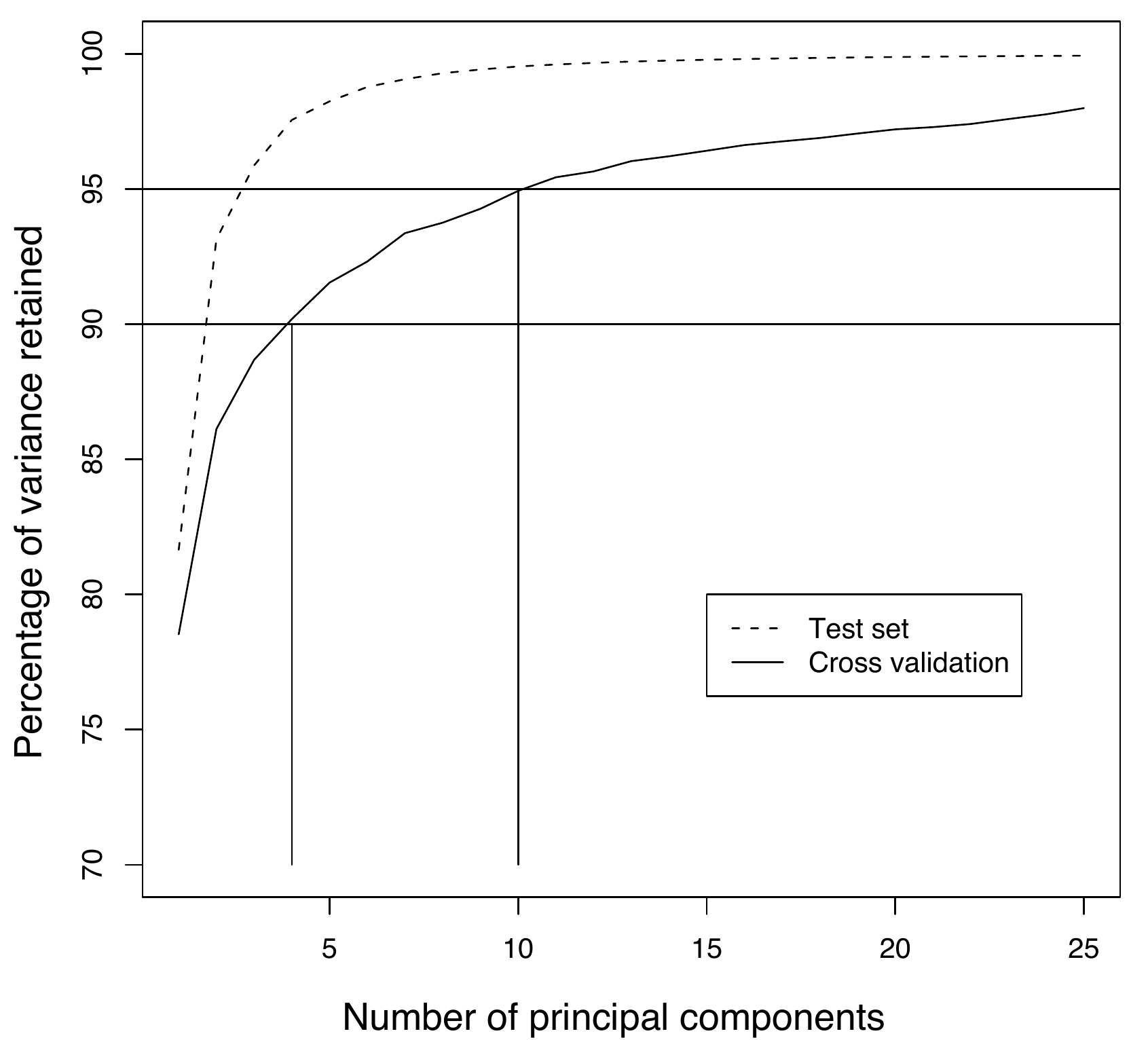}
\caption{Capacity of principal components to retain the measurement variance.}
\label{PCAVarRet}
\end{figure}

Fig. \ref{PCAVarRet} provides the average retained variance on the 10 test sets for the first 25 principal components. The upper line gives the average amount of variance retained when the principal components are computed with the test sets, and provides an upper bound on the compression efficiency that the PCA can achieve on this dataset. The lower curve gives the average amount of variance retained on the test set when the components are computed with the training set. This figure shows that the first principal component accounts on average for almost $80\%$ of the variance, while $90\%$ and $95\%$ of variance are retained with $4$ and $10$ components, respectively. The confidence level of these estimates was about $\pm 5\%$. Additional experiments were run using $K$-fold cross validation with $K$ ranging from $2$ to $30$. The percentages of retained variance on the test data blocks tended to decrease with $K$. Losses of a few percents were observed for $K$ higher than $15$ (less than nine hours of data). 

The amount of retained variance increases very fast with the first principal component, and becomes almost linear after about ten components. A linear increase of retained variance with the number of principal components reflects the fact that the components obtained by the PCA are actually no better than random components \cite{jolliffe2002pca}. From Fig. \ref{PCAVarRet}, it therefore seems that from $10$ or $15$ components onwards, the remaining variations can be considered as white noise. 

\begin{figure}[!ht]
    \centering
      \includegraphics[width=10cm]{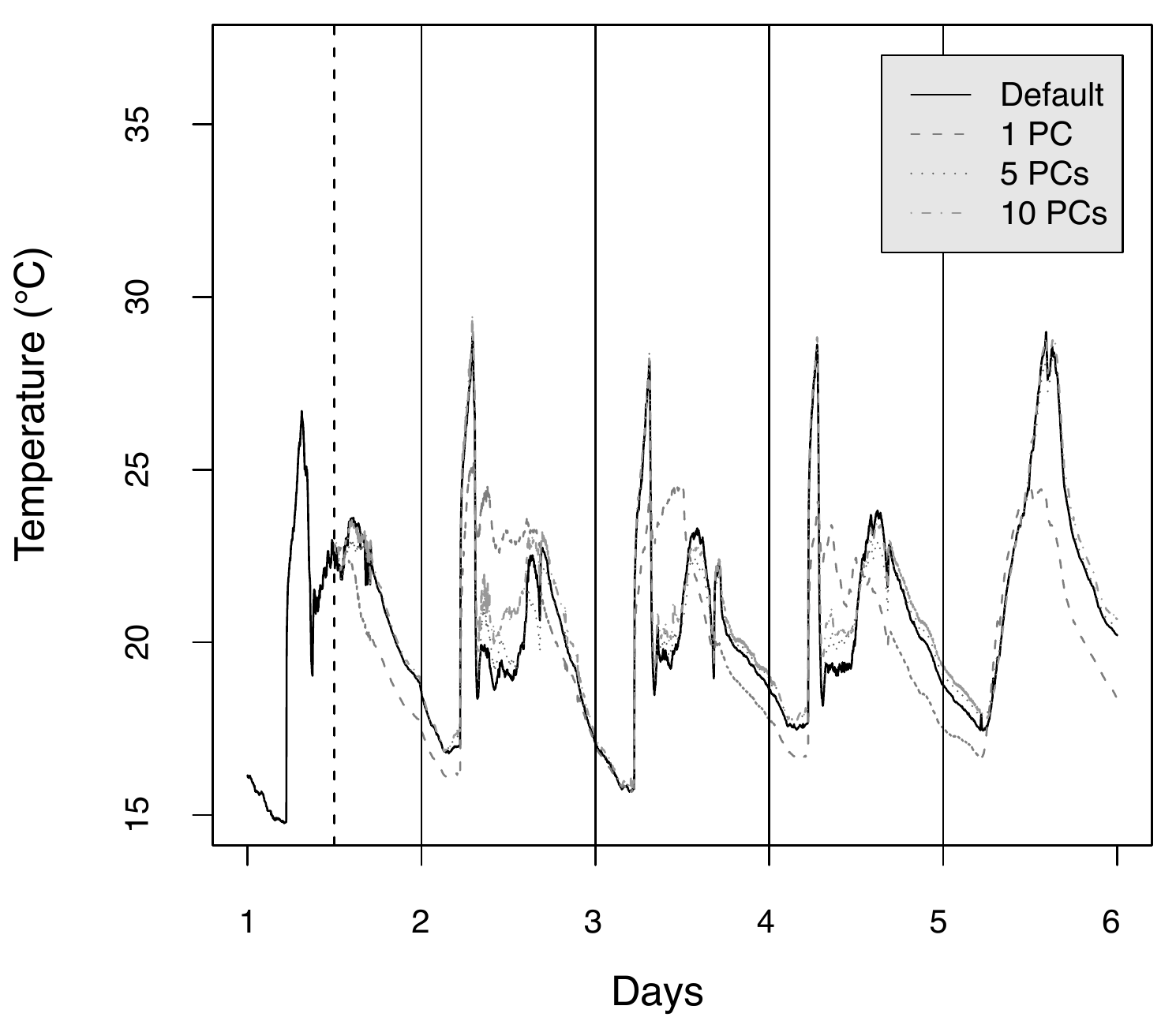}
\caption{Approximations obtained on the test set for the sensor 21, using one, five and fifteen principal components. The covariance matrix was computed from the first twelve hours of measurements (before the vertical dashed line).}
\label{figApprox49Intel}
\end{figure}

Fig. \ref{figApprox49Intel} illustrates the approximations obtained during the first round of the cross validation (i.e., principal components are computed from the first 12 hours of measurements) for the sensor 49, using one, five and ten principal components. A single principal component provides rough approximations, which cannot account for fine-grained details of some of the sensor measurements. For example, the stabilization of the temperature around $20^\circ {\rm C}$ around noon during the second, third and fourth day (probably due to the activation of an air conditioning system at a location close to sensor 49) are not captured by the approximations. Increasing the number of principal components allows to better approximate the local variations, and passing to five components provides for example a much better approximation of the sensor measurements. 

\subsection{Communication costs}
\label{respcaggnet}

\begin{figure}[!ht]
    \centering
      \includegraphics[width=18cm]{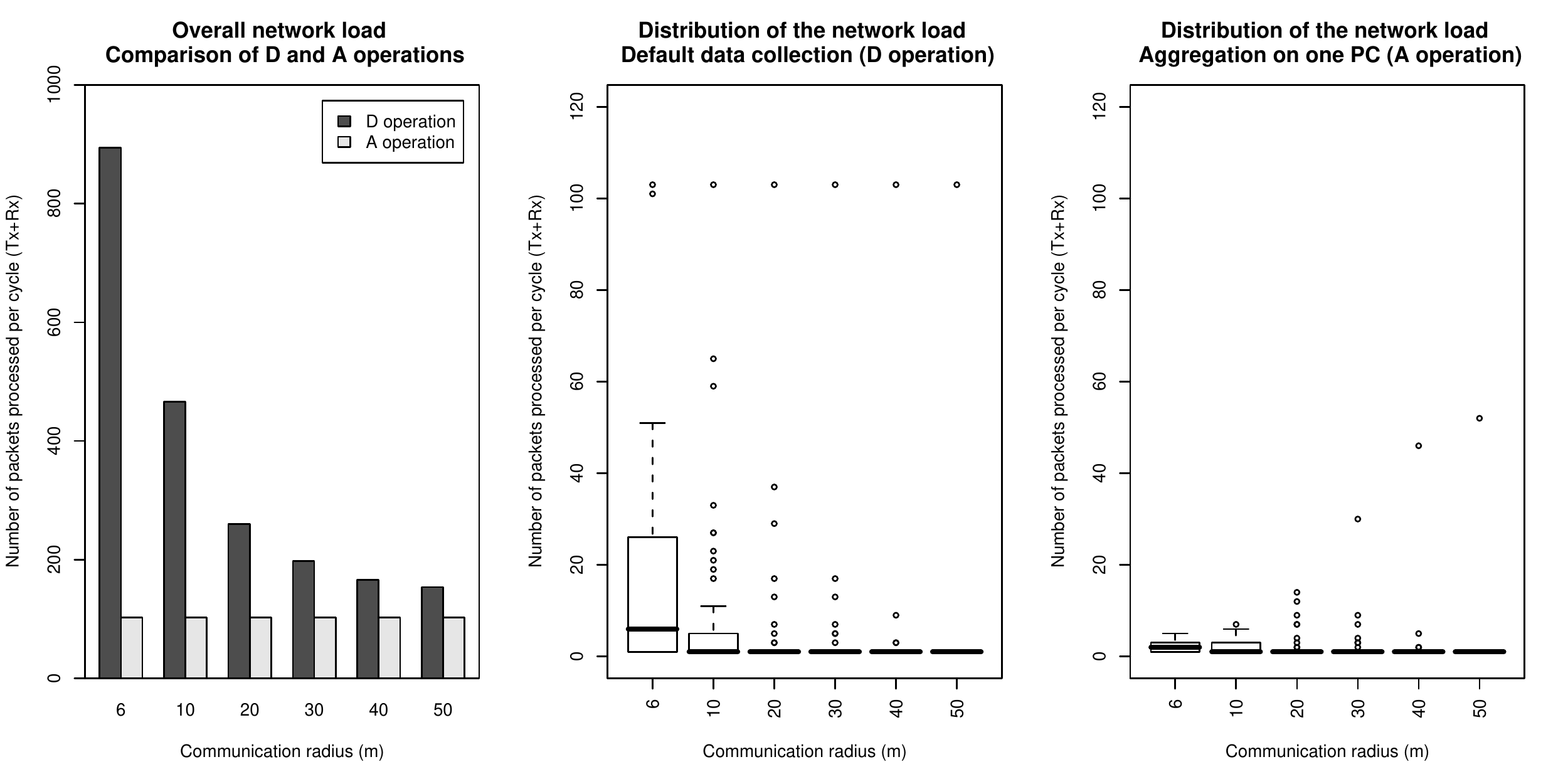}
\caption{Communication costs entailed by D and A operations. The overall network load (left) is in all cases higher for a D operation than for an A operation. Details of the distribution of the network load per node reported as boxplots (middle and right) reveal that the distributions of the load are more balanced using aggregation. Aggregation also leads to reduce the highest network load.}
\label{IntelCommCost2}
\end{figure}

We compare in the following the communication costs entailed by the default and PCAg schemes for different types of routing trees. Fig. \ref{IntelCommCost2} reports the total number of packets processed by nodes (received and transmitted) during an epoch, for the default and PCAg scheme, as the radio communication range increases. The subfigure on the left reports the overall sensor network load, while subfigures in the middle and on the right detail with boxplots the distributions of the network load per node in the network, for the default and PCAg schemes, respectively. 

Given that sensors choose as their parent the sensor within radio range that is the closest to the base station, increasing the radio communication range typically leads the routing tree to have a smaller depth, and its nodes to have a higher number of children. As was pointed out in Section \ref{tradeoffPCA}, this is detrimental to the PCAg scheme, as this may lead the numbers of packets received by a node to be higher than in the default scheme. 

For the default scheme, increasing the radio range reduces the overall sensor network load (Fig. \ref{IntelCommCost2}, left), and eventually leads all the nodes but the root to only transmit one packet (Fig. \ref{IntelCommCost2}, middle, radio range of 50m). In this extreme case, the routing tree has depth one, and all nodes but the root are leaf nodes. Note that the highest network load does not depend on the tree topology. This highest load is sustained by the root node which, whatever the depth of the tree, is required to forward other node's packets (i.e., 51 measurements to receive and send), and to send its own measurement to send. Its network load is therefore of 103 packets per epoch.

For the PCAg scheme, we first report results for the extraction of one component. A nice feature of aggregation is that the overall network load (Fig. \ref{IntelCommCost2}, left) does not depend on the topology, thanks to the fact that forwarding does not increase the number of transmissions. Looking at the details of the distribution of the network load per node (Fig. \ref{IntelCommCost2}, right), it is interesting to see that increasing the radio range has a reverse effect for the PCAg scheme, as it tends to increase the network load. This is a direct consequence of the increased number of children induced by routing trees with smaller depths. Eventually, for a fully interconnected network, we observe the same effect than for the default scheme, where all the sensors send only one packet, while the root node sustains the higher network load due to the forwarding task. Note however than thanks to the aggregation process, it only sends one packet, which reduces to $52$ packets per epoch its network load (one packet transmitted and $51$ packets received). The extraction of one component therefore decreases the network load supported by the root node. The network load incurred by $k$ components is obtained by multiplying the values found in Fig. \ref{IntelCommCost2} by $k$. The PCAg scheme may therefore be less efficient than the the default scheme if many components are extracted.

\begin{figure}[!ht]
    \centering
      \includegraphics[width=10cm]{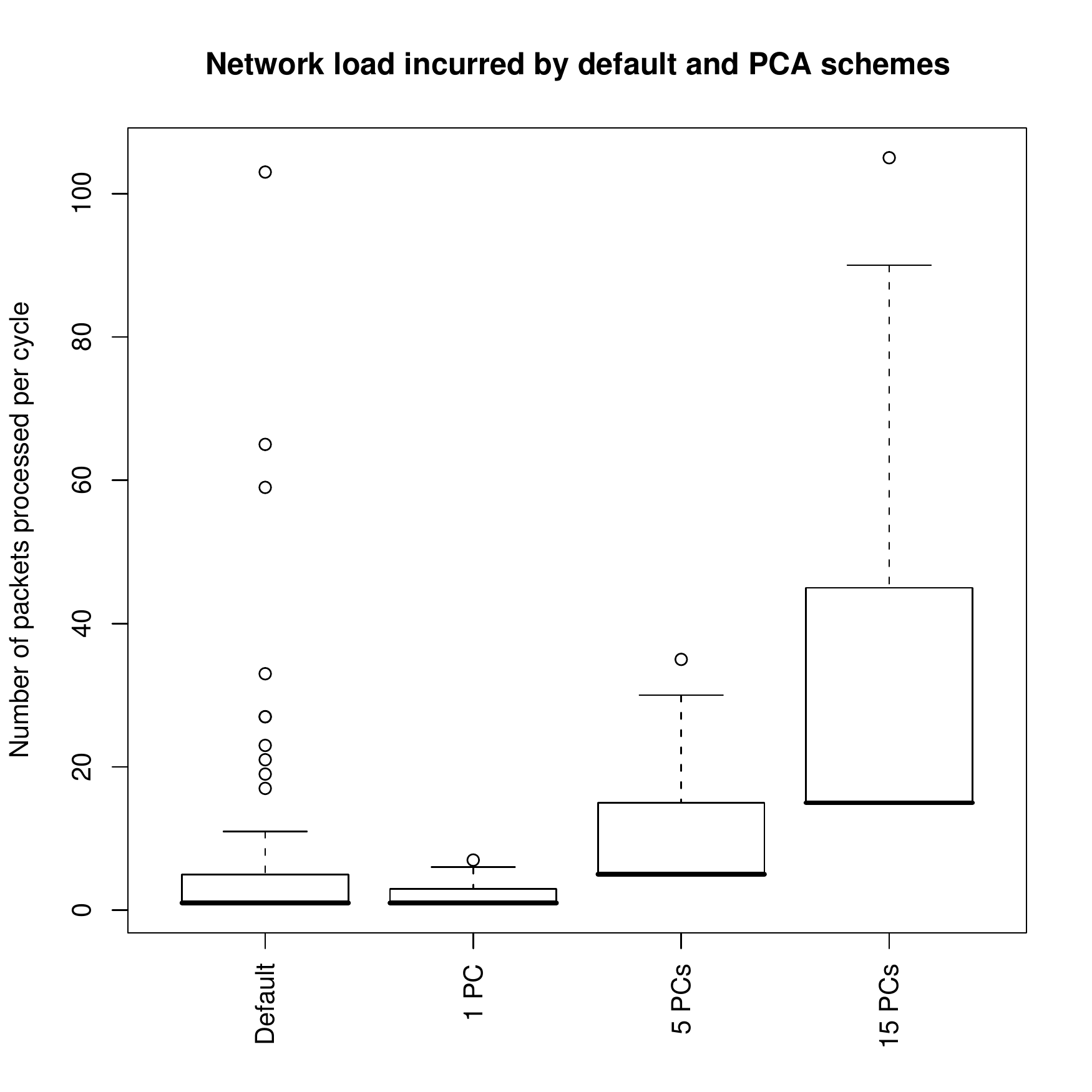}
\caption{Comparison of the communication costs entailed by a D operation, and A operations with $1$, $5$ and $15$ principal components. Radio range is 10 meters.}
\label{IntelCommCost1}
\end{figure}

This is illustrated in Fig. \ref{IntelCommCost1} where the number of packets processed (received and sent) is reported as a function of the number of principal components extracted for a radio range of $10$ (Illustrated in Fig. \ref{routingTree}). In this routing tree, the highest number of children is $6$. For the extraction of one PC, the highest network load is therefore of $7$, i.e., $6$ receptions and one transmission, to be compared with the highest network load of $103$ for the root node in the default scheme. This results in a reduction of about $85\%$ of the network load. Extracting more than $15$ components leads however the highest network load to be higher than in the default scheme, as the sensor node aggregating the packets from its 6 children will sustain a network load of $105$ packets per epoch.

The overall network load generated by the aggregation of $k$ components is in the same manner $k$ times the load generated by of the aggregation of one component. For a communication radius of $10$ meters, this overall load was of $103$ packets as reported in Fig. \ref{IntelCommCost2}, left, to be compared with an overall load of $466$ packets for the default scheme.  The overall load generated by the aggregation process is therefore higher than the load of the default scheme from $5$ components. This is due to the fact that aggregation, while decreasing the load of the most solicited node, increases evenly the load of all other nodes.

\subsection{Distributed covariance matrix}
\label{sup}

\begin{figure}[!ht]
\begin{minipage}{.45\textwidth}
    \centering
    \vspace{0.3cm}
      \includegraphics[height=6cm]{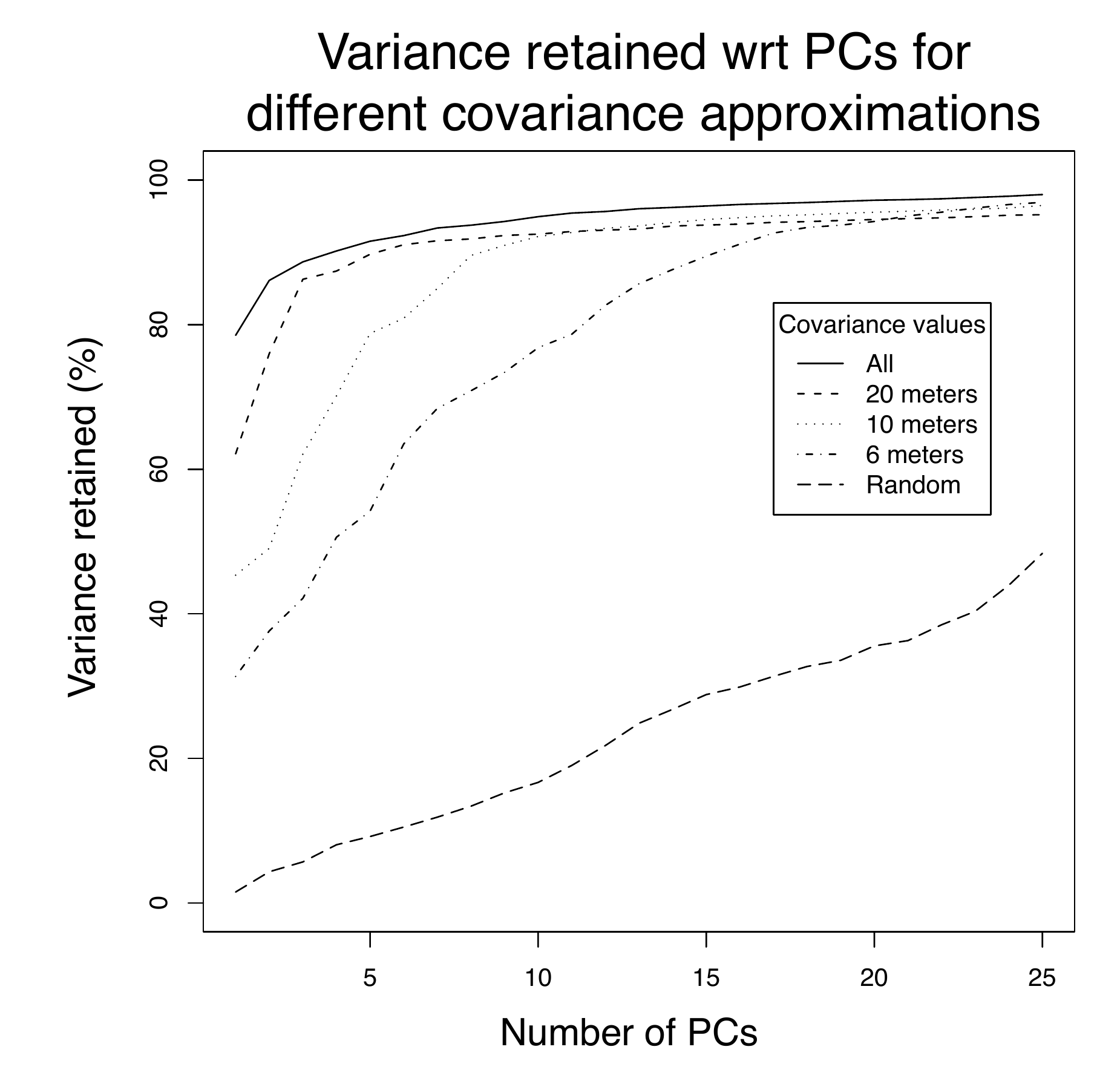}
\caption{Capacity of the principal components to retain the measurement variance.}
\label{IntelVarRetCrop}
\end{minipage}
\hfill
\begin{minipage}{.45\textwidth}
     \centering
      \includegraphics[width=6.2cm]{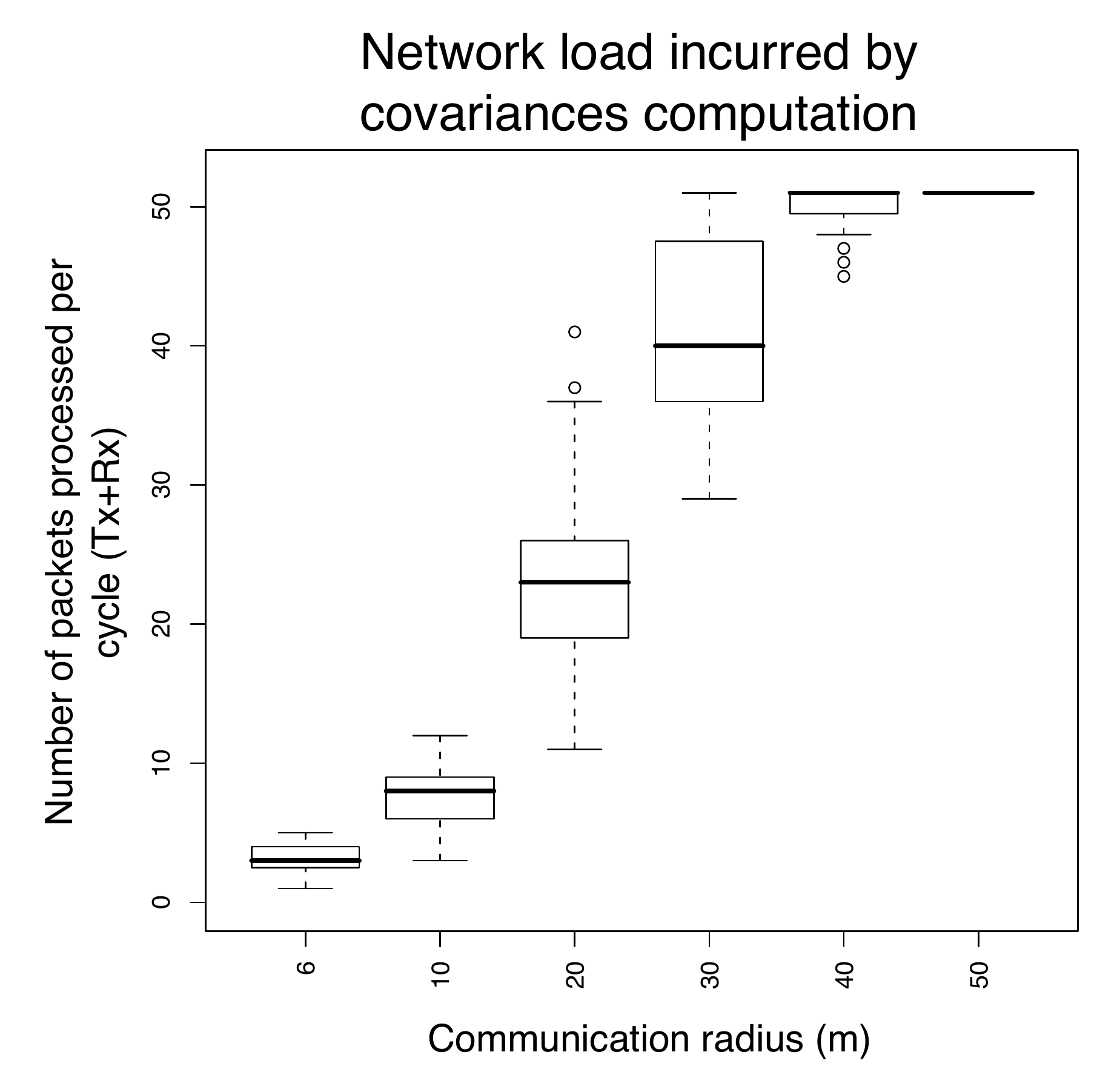}
\caption{Network load incurred by local covariances updates.}
\label{IntelCovMatNetCost}
\end{minipage}
\end{figure}

We study in this section the ability of the local covariance hypothesis to properly identify the principal component subspace, and discuss how the accuracy loss is counterbalanced by gains in energy consumption and network load. In Fig. \ref{IntelVarRetCrop}, the upper curve gives the amount of variance retained if all covariances are computed, and is the same as in Fig. \ref{PCAVarRet}. Lower curves correspond to the percentage of variance retained as the radio range of sensors is decreased, and illustrate the fact that the local covariance hypothesis may have as a negative consequence a loss of accuracy for identifying the principal components. If the local covariance hypothesis does not hold, the loss can be high, as is illustrated by a radio range of $6$ meters. This loss is however attenuated when the number of principal components increases. In any case, the subspace obtained by computing the principal components from the approximate covariance matrix is significantly better in retaining information than a random subspace.

We note that reducing the radio range also decreases the energy spent in communication, as a lower transmission power leads to a lower energy consumption for the radio. We illustrate in Fig. \ref{IntelCovMatNetCost} the potential savings in terms of network load obtained by decreasing the radio range. This figure reports the distribution of the network loads related to the computation of the covariance matrix for varying radio ranges. The average network load increases with the radio range, as the neighborhood of sensors gets larger. Fig.  \ref{IntelVarRetCrop} and \ref{IntelCovMatNetCost} provide together an illustration of the tradeoff between the accuracy and the communication costs entailed by the local covariance hypothesis. A radio range range larger than $20$ meters does not bring much gains in terms of accuracy, while it strongly increases the network load.  

Also, it is interesting to compare the results reported in Fig. \ref{IntelCovMatNetCost} and Fig. \ref{IntelCommCost2}. The highest network load caused by an update is in all cases lower using the distributed covariance matrix scheme ($52$ packets processed against $101$ at the root for the default data gathering operation). The average network load can however be higher as the radio range increases, due to the fact that nodes close to the leaves process more packets in the distributed scheme than in the centralized scheme.

\subsection{Distributed principal component computation}
\label{resdpca}

\begin{figure}[!ht]
\begin{minipage}{.45\textwidth}
    \centering
      \includegraphics[width=6cm]{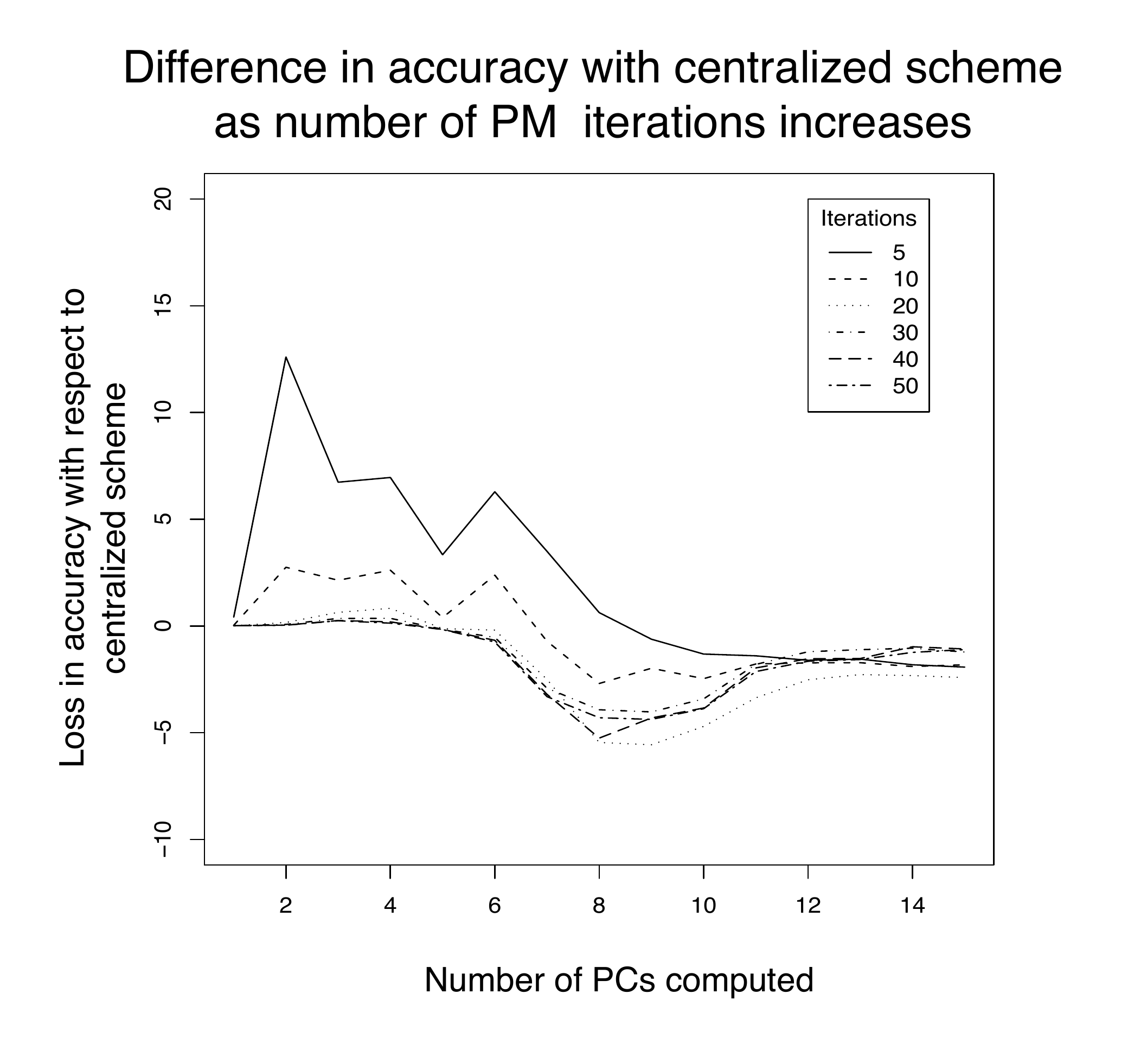}
\caption{Comparison of the accuracy obtained using the exact and the approximated eigenvectors, for different number of iterations and principal components.}
\label{PMAccNbIter}
\end{minipage} 
\hfill
\begin{minipage}{.45\textwidth}
    \centering
        \vspace{-0.6cm}

      \includegraphics[width=6cm]{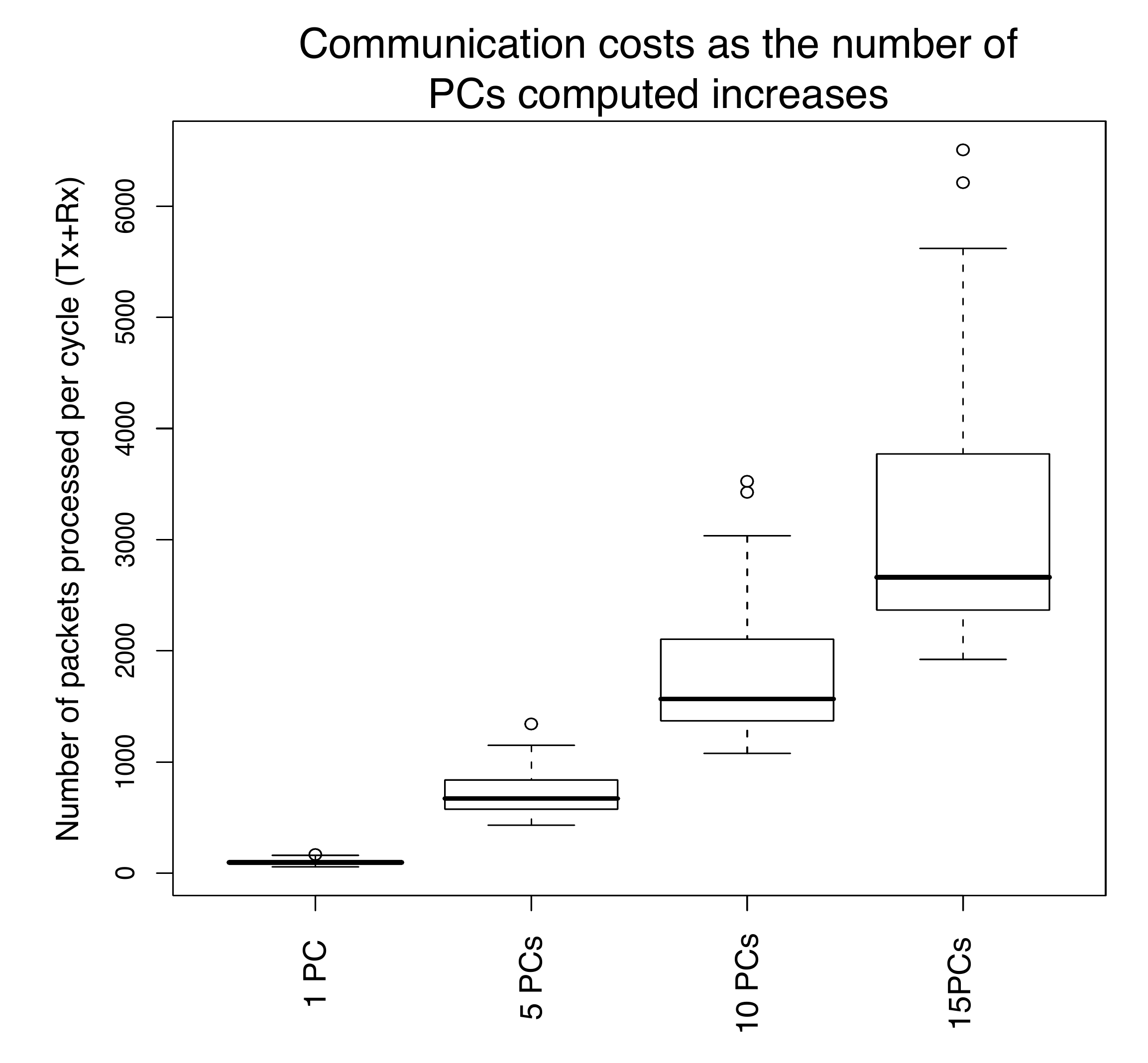}
\caption{The network load is quadratic in the number of principal components. Radio range is 10 meters.}
\label{PMNetCosts}
\end{minipage}
\end{figure}

We finally discuss  the ability of the power iteration method to properly identify the principal components. The main parameter for the estimation of the eigenvectors is the convergence criteria used. We illustrate in Fig. \ref{PMAccNbIter} the difference in accuracy obtained on the test set, for different convergence criteria, between the set of exact eigenvectors (Computed in a centralized manner with the QR method), and the set of approximated eigenvectors obtained by means of the power iteration method. The convergence threshold $\delta$ was set to $10^{-3}$, and we tested the accuracy obtained for $5$, $10$, $20$, $30$, $40$ and $50$ iterations. Results reported are averages of ten-fold cross validations, and the confidence level of the results was about $2\%$. 

This Figure allows to experimentally illustrate a set of possible behaviors, which are understood by keeping in mind that the rate of convergence of the power method depends on the ratio of the two dominant eigenvalues. Typically, in correlated data, the ratio of subsequent eigenvalues decreases exponentially (which can be seen qualitatively in Fig. \ref{PCAVarRet}), making the convergence speed lower as the number of principal components computed increases. As a result, few iterations are usually enough to converge to the first eigenvectors. This is seen in Fig.  \ref{PMAccNbIter} for the first PC, for which as little as five iterations are enough to nicely converge. In the computation of the subsequent PCs, we observe that the number of iterations required to properly converge is higher (about 20 iterations led to accuracy similar to the centralized approach). If the number of iterations is not high enough, the power iteration leads to a PC that may not represent the data as well as the centralized approach, as is observed from the second PC for a maximum number of five iterations. Note that, given the fact that PCs are estimated on the basis of past measurements whose distribution is not exactly the same as upcoming measurements, it may also happen that approximations provide better accuracies than the centralized approach. This explains the gains in accuracy obtained around the $8$-th component where the difference between subsequent eigenvalues gets very low. 

Additional experiments were also run to observe the consequence of the local covariance hypothesis on the non positive semi-definiteness of the matrix. We observed that negative eigenvalues could lead the algorithm to stop at stages as early as between $5$ to $10$ components (this was observed for radio ranges of $30$ and $40$ meters). It is however important to note that despite the early stopping of the algorithm, the set of principal components identified was enough to retain more than $90\%$ of the variance (cf. Fig. \ref{IntelVarRetCrop}). 

Finally, we report in Fig. \ref{PMNetCosts} estimates of the communication costs entailed by the computation of the principal components. The radio range was of $10$ meters, and the costs were computed following the analysis of Sections \ref{comcosts} and \ref{summaryscheme}. While about two hundred packets per node suffice for the computation of the first eigenvector, this quantity can reach a significant network load of $6000$ packets on average for the computation of $15$ components. As discussed in Section \ref{summaryscheme}, the main advantage of the distributed computation of the principal components lies in the distribution of the computational and memory costs, and the approach does not scale well to the computation of a large number of principal components.

\subsection{Summary}
\label{resdisc}

The main parameter of the principal component aggregation, which determines the tradeoff between the compression accuracy and the communication costs, is the number of principal components to extract. As illustrated in Fig. \ref{IntelCommCost1} or \ref{PMNetCosts}, the network load incurred by the computation of the principal components quickly increases with their number. The PCAg scheme therefore proves useful only if a few components are required. In such a case, we emphasize that the gains can be dramatic, as was reported in Fig.  \ref{IntelCommCost1} or \ref{PMNetCosts}. 

The amount of information retained by a set of principal components depends on the correlation existing between sensor measurements. We illustrated the method for a compression task using a real-world temperature dataset where measurements had an degree of correlation that can be assumed to be representative of a typical sensor network scenario. An interesting result to recall is that for $5$ principal components, $90\%$ of the variance could be retained. This was shown to return a nice approximation of the original signals (Fig. \ref{figApprox49Intel}), and to reduce the highest network load from about $100$ packets per epoch to about $40$ packets per epoch (Fig. \ref{IntelCommCost1}). 

Regarding the scalability, an interesting property of the PCAg is that the network load is better distributed in the network. This property first prevents congestion issues observed at the root node using the default scheme. Second, as the radio communication is a primary source of energy depletion for a wireless node, it better distributes the lifetime of the nodes in the network. The PCAg therefore provides an interesting framework for dealing with data in multi-hop networks.

\section{Related work}
\label{section5}

The distribution of the PCA has been addressed in the signal processing and information theory communities \cite{vosoughi_s2007padpfddc_article,pradhan_r2003dscusddac_article,xiong2004dsc,Gastpar2006dklt}. These approaches, based on distributed source coding, involve compression techniques related to Slepian-Wolf and Wyner-Ziv coding together with Karhunen-Loeve transform for compressing data in distributed systems. Each component of the system is however assumed to observe several dimensions of the problem, and they require the base station to support the task of defining the coding scheme. The network architectures and communication costs analyses are also left as open research areas. Although these approaches may lead to attractive applications for WSN, they still remain in that respect at an early research stage.

In the fields of machine learning and data mining, distributed PCA schemes have also been addressed in \cite{Bai2005pcad,kargupta2000cpc} and \cite{huang2006npa}. The former approaches aimed at combining the PCA results obtained from different sources that share the same variables (\emph{vertically} distributed databases). The latter work provided an architecture aimed at network anomaly detection, where the PCA is first computed in a centralized manner, and a subsequent distributed and adaptive feedback mechanism is proposed. Despite the similarity in the choice for the method name, the corresponding work is however clearly different from the approach proposed in this article.

The distributed computation of eigenvectors of large and sparse matrices has been tackled from several angles in the literature, and good reviews of existing state of the art techniques are for example detailed in \cite{bai2000tsa,golub1996mc}. The approach proposed in this article is however assumed to be innovative to the best of authors' knowledge, as it leverages two specific architectural properties of sensor networks. First, a wireless sensor network is an \emph{intrinsically} distributed system where each component only captures one dimension of the system variations. Second, the communication constraints can be coupled with the local covariance hypothesis. 

Recent work in the domain of link analysis and recommendation systems has led authors in \cite{jelasity2007adp} to propose a distributed implementation of the power iteration method, which closely matches the approach proposed in this article. Their algorithm aims at computing the principal eigenvector of the adjacency matrix of a peer-to-peer network, which leads to the ranking of the network components, in a way similar to the \emph{page rank} algorithm. The underlying hypotheses of the distributed system are interestingly close to ours (each component has only access to one dimension the problem, and can communicate with components whose data is related). The network structure is however different, as no coordinating entity is assumed. This leads authors to rely on the harmonic mean to achieve the normalization step of the power iteration method. The computation of subsequent eigenvectors is also not addressed. 

Related work on the use of basis change for sensor networks has been addressed in \cite{duarte2005dcs2}, where authors proposed the radical position of using random bases to project sensor measurements. The work is analyzed in the context of compressed sensing, an emerging field in signal processing (see also \cite{donoho2006cs}). Their work has however mainly focused on the theoretical ability of random bases to retain the sensor measurements variations.

\section*{Conclusion}

This article extended previous work related to the distribution of the principal component analysis by presenting an implementation suitable for wireless sensor networks. The approach relies on the hypothesis that sensor measurements collected by distant sensors are uncorrelated, and was shown to provide significant gains in terms of radio communication. Additionally, we showed that the distributed principal component analysis led to balance the network load among the sensors, making the method particularly suitable for multi-hop sensor networks. We plan to extend this work by showing that spatiotemporal aggregation and independent component analysis can also be formulated in the same framework. 

\section*{Acknowledgments}
This work was supported by the COMP2SYS project, sponsored by the Human Resources and Mobility program of the European Community (MEST-CT-2004-505079), and by the PIMAN project, supported by the Institute for the Encouragement of Scientific Research and Innovation of Brussels, Belgium.

\bibliographystyle{sensors}
\bibliography{biblio}

\vspace{12pt}\noindent \copyright \ 2008 by the authors; licensee Molecular Diversity Preservation International, Basel, Switzerland.
This article is an open-access article distributed under the terms and conditions of the Creative
Commons Attribution license (http://creativecommons.org/licenses/by/3.0/).

\end{document}